\documentclass[prb,aps,twocolumn,groupedaddress,nofootinbib,10pt]{revtex4-2} 
\usepackage{epsfig}
\usepackage{amsfonts}
\usepackage{amscd}
\usepackage{latexsym}
\usepackage{amsmath,amssymb}
\usepackage{verbatim}
\usepackage{setspace}
\usepackage[dvipsnames]{xcolor}
\usepackage{fancyhdr}
\usepackage{hyperref}
\usepackage{slashed}
\usepackage{multirow}
\usepackage{chiraldraft}
\usepackage{mciteplus}
\usepackage{bbm}
\usepackage[english]{babel}
\usepackage{amsthm}
\usepackage{soul}
\newcommand{\QV}{Q^{\rm V}}
\newcommand{\QA}{Q^{\rm A}}

\newcommand{\cQA}{\cQ^{\rm A}}
\newcommand{\tQA}{\t{Q}^{\rm A}}
\newcommand{\qV}{q^{\rm V}}
\newcommand{\qA}{q^{\rm A}}
\newcommand{\tqA}{\tilde q^{\rm A}}
\newcommand{\QVA}{Q_{\rm com}}

\newcommand{\hc}{{\rm h.c.}}

\renewcommand{\ie}{\begin{equation}\begin{aligned}[]}

\renewcommand{\theequation}{\arabic{equation}}

\hypersetup{
colorlinks,citecolor=pastelblue,linkcolor=pastelblue,urlcolor=pastelblue}

\usepackage{color}
\definecolor{red}{rgb}{1,0,0}
\definecolor{blue}{rgb}{0,0,1}
\definecolor{dblue}{rgb}{0,0,0.4}
\definecolor{green}{rgb}{0,1,0}
\definecolor{black}{rgb}{0,0,0}
\definecolor{white}{rgb}{1,1,1}
\definecolor{pastelblue}{RGB}{20,93,160}

\definecolor{brn}{rgb}{.8,.4,.0}
\definecolor{redo}{rgb}{1,.5,.0}
\definecolor{ddgrn}{rgb}{0,0.4,0}
\definecolor{dgrn}{rgb}{0,0.55,0}
\definecolor{dbl}{rgb}{0,0,0.5}

\newcommand{\Z}{\mathbb{Z}}

\renewcommand{\t}[1]{\widetilde{#1}} 

	\newcommand{\cO}{ {\cal O} } 

\newcommand{\al}{\alpha} 
	\newcommand{\bt}{\beta} 
	\newcommand{\del}{\delta}

	\newcommand{\ga}{\gamma}

	\newcommand{\lm}{\lambda} 
	 
	\newcommand{\om}{\omega} 
	\newcommand{\Om}{\Omega}

\usepackage{comment}
\usepackage{hhline}

\newcommand*{\RightThumbsUpAux}[1]{%
  \begingroup
    \sbox0{Ag}%
    \raisebox{-\dp0}{%
      \includegraphics[{%
        height=\dimexpr\dp0+\ht0\relax,
        #1%
      }]{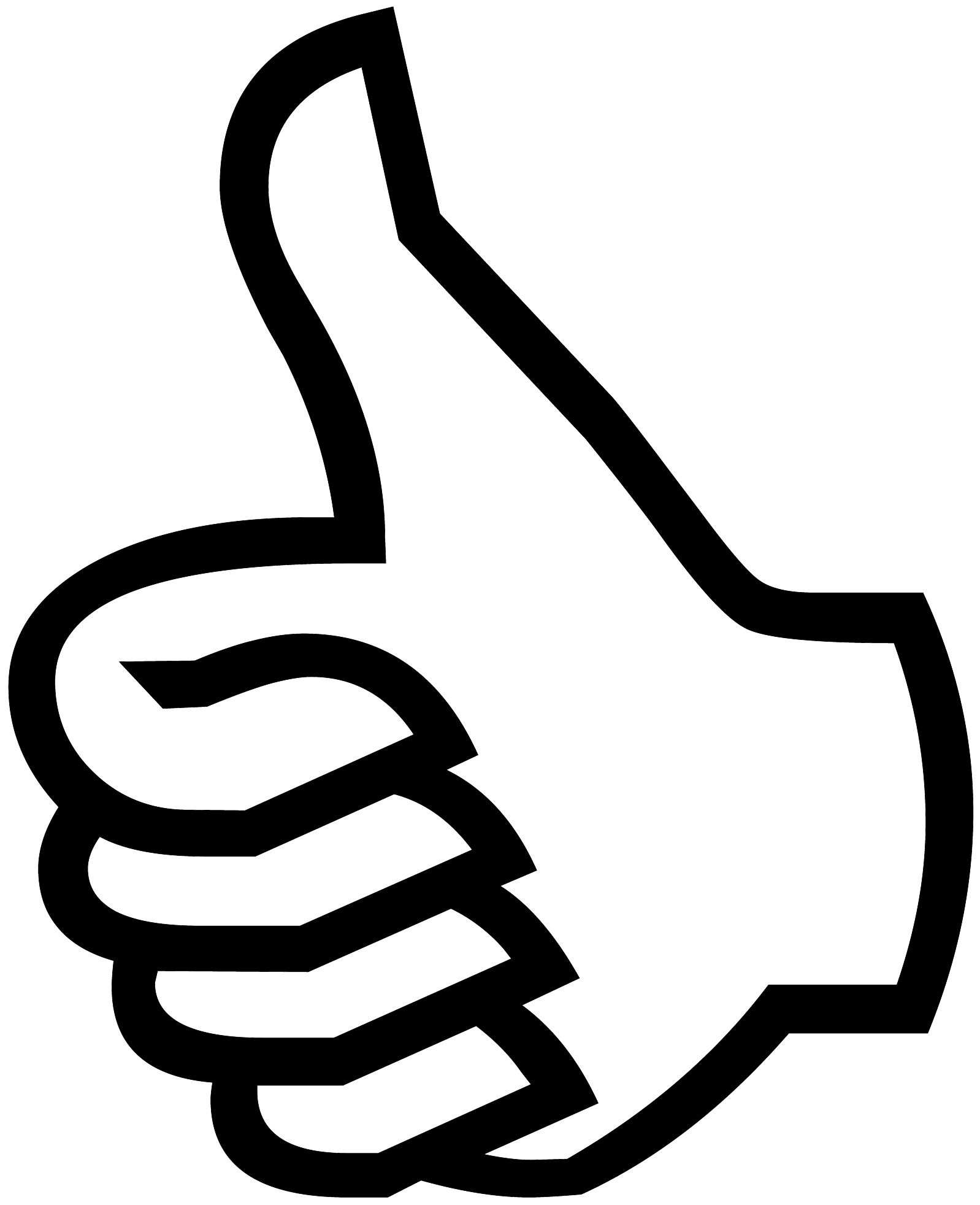}%
    }%
  \endgroup
}
\newcommand*{\RightThumbsUp}{%
  \RightThumbsUpAux{}%
}

\newcommand*{\LeftThumbsUp}{%
  \scalebox{-1}[1]{\RightThumbsUp}%
}

\begin{document}

\hfill MIT-CTP/5760 \quad YITP-SB-2024-19

\title{Quantized axial charge of staggered fermions and the chiral anomaly}

\author{Arkya Chatterjee$^{\,\text{\footnotesize\LeftThumbsUp}}$, Salvatore D. Pace$^{\,\text{\footnotesize\LeftThumbsUp}}$, Shu-Heng Shao$^{\,\text{{\footnotesize\LeftThumbsUp},{\footnotesize\RightThumbsUp}}}$}
\affiliation{$^{\text{\footnotesize\LeftThumbsUp}}$Department of Physics, Massachusetts Institute of Technology,
 Cambridge, Massachusetts 02139, USA
 \\
$^{\text{\footnotesize\RightThumbsUp}}$Yang Institute for Theoretical Physics, Stony Brook University, Stony Brook, New York 11794, USA}

\begin{abstract}
In the 1+1D ultra-local lattice Hamiltonian for staggered fermions with a finite-dimensional Hilbert space, there are two conserved, integer-valued charges that flow in the continuum limit to the vector and axial charges of a massless Dirac fermion with a perturbative anomaly. 
Each of the two lattice charges generates an ordinary U(1) global symmetry that acts locally on operators and can be gauged individually. 
Interestingly, they do not commute on a finite lattice and generate the Onsager algebra, but their commutator goes to zero in the continuum limit. 
The chiral anomaly is matched by this non-abelian algebra, which is consistent with the Nielsen-Ninomiya theorem. 
We further prove that the presence of these two conserved lattice charges forces the low-energy phase to be gapless, reminiscent of the consequence from perturbative anomalies of continuous global symmetries in continuum field theory. Upon bosonization, these two charges lead to two exact U(1) symmetries in the XX model that flow to the momentum and winding symmetries in the free boson conformal field theory.

\end{abstract}

\maketitle

\tableofcontents

\section{Introduction}
The realization of chiral symmetries on the lattice has been a long-standing problem in lattice field theory~\cite{Kaplan:2009yg}. 
One prominent difficulty is that the global symmetries of interest typically have anomalies, in the sense that they cannot be coupled to dynamical gauge fields in a consistent manner.\footnote{Anomalies of global symmetries are often called 't Hooft anomalies, which indicate that even though the global symmetries are conserved and exact at the quantum level, there is an obstruction to gauging them. This contrasts with the Schwinger anomaly~\cite{PhysRev.128.2425} and the Adler-Bell-Jackiw (ABJ) anomaly~\cite{Adler:1969gk,Bell:1969ts}, where a classical symmetry fails to be a true global symmetry at the quantum level. See, for example, Ref.~\cite{Kapustin:2014lwa} for this terminology. These two notions of anomalies are not unrelated. For example, the Schwinger anomaly of the axial symmetry in 1+1D QED arises from the 't Hooft anomaly involving both the axial and vector symmetries of a free Dirac fermion. The violation of the axial symmetry when the vector symmetry is gauged implies an obstruction to gauging both global symmetries simultaneously.}

Anomalies are traditionally linked to divergences in continuum field theory and are 
 often considered challenging, if not impossible, to realize on a lattice with finite lattice spacing. 
However, this piece of lore is not true. 
Numerous examples of anomalies for discrete global symmetries are realized on the lattice. See, for example, Ref.~\cite{Seifnashri:2023dpa} for a recent survey of this topic.

A more refined piece of lore is that perturbative anomalies of continuous global symmetries cannot be realized on a lattice. 
By perturbative or local anomalies, we mean those captured by Feynman diagrams and local operator product expansions.
This piece of lore also turns out to be incorrect: perturbative anomalies can be realized in Villain-type lattice models~\cite{Sulejmanpasic:2019ytl,Gorantla:2021svj,Cheng:2022sgb,Fazza:2022fss,Seifnashri:2023dpa,Berkowitz:2023pnz}. 
The local Hilbert spaces of these models involve continuous variables and are infinite-dimensional. 
See Ref.~\cite{Berkowitz:2023pnz} for discussions of chiral fermion symmetry in 1+1D QED in this setup, and Ref.\ \cite{Catterall:2018lkj} for a Euclidean lattice realization of mixed gravitational anomalies with global symmetries.

Perhaps a more interesting question is: \textit{can perturbative anomalies be realized in a lattice model whose local Hilbert space is finite-dimensional (e.g., qubits)?} 
There is a simple no-go argument against this hope. 
This can be seen most readily for U(1) symmetries in 1+1D, 
where perturbative anomalies are encoded in the equal-time commutator of the conserved current operators ${[j^0(t,x) ,j^x (t, x')] \sim i \partial_x\delta( x- x')}$, known as the Schwinger term~\cite{Schwinger:1959xd,Treiman:1986ep}.\footnote{We thank Tom Banks and Nathan Seiberg for this discussion.} 
Taking the trace of this relation immediately shows that the Schwinger term cannot be realized verbatim on a finite-dimensional Hilbert space since the trace of a commutator in a finite-dimensional vector space necessarily vanishes. 
This is analogous to the fact that ${[X,P]=i\hbar}$ cannot be realized in a finite-dimensional Hilbert space in quantum mechanics. 
See Ref.~\cite{Kapustin:2024rrm} for a rigorous proof of this no-go result.
The Schwinger term, however, has recently been realized exactly on a lattice model with an infinite-dimensional Hilbert space~\cite{Cheng:2022sgb}.

\begin{figure}
    \centering
\includegraphics[width=0.95\linewidth]{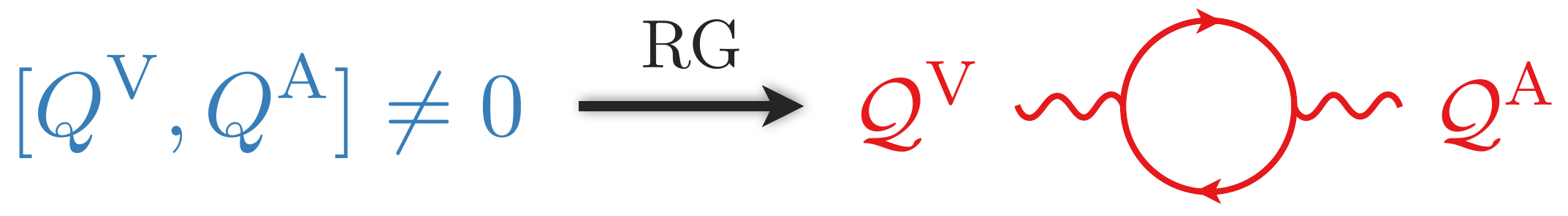}
    \caption{In an ultra-local Hamiltonian lattice model, we discuss two conserved, quantized charges $\QV$ and $\QA$, which become the vector and axial charges $\mathcal{Q}^\text{V}$ and $\mathcal{Q}^\text{A}$ of the massless Dirac fermion field theory in the continuum limit. We use calligraphic and ordinary fonts for operators in the continuum and on the lattice, respectively. The anomaly between ${\cal Q}^\text{V}$ and ${\cal Q}^\text{A}$ is matched on the lattice by the non-abelian Onsager algebra generated by $\QV$ and $\QA$.}
    \label{fig:QVQA}
\end{figure}

These difficulties are rigorously formulated in the Nielsen-Ninomiya theorem \cite{Nielsen:1980rz,Nielsen:1981hk,Nielsen:1981xu}. 
The theorem, most precisely formulated in~\cite{Friedan:1982nk}, asserts that in any quadratic lattice fermion Hamiltonian with certain locality properties in odd spatial dimensions, there must be an equal number of left- and right-moving fermions within each irreducible representation of the global symmetries.\footnote{This statement is only meaningful once a specific charge sector of the global symmetries is fixed. Otherwise, for instance, complex conjugation changes the chirality and global symmetry charge of a Weyl fermion field in 3+1D.}
In particular, this theorem prohibits the existence of an axial charge that (i) has quantized, integer eigenvalues and (ii) commutes with the vector U(1) symmetry. If such an axial charge existed, one could restrict to the fixed charge sector of both the vector and axial symmetries and would find a single left-moving fermion.

Given the above, what fingerprints of anomalies can we hope for on a lattice with a finite-dimensional Hilbert space? 
In this letter, we focus on the anomaly between the vector and axial symmetries of a 1+1D massless Dirac fermion, which is the oldest and arguably the simplest anomaly of all. 
In a microscopic Hamiltonian lattice realization, we discuss two conserved, quantized charges that become the vector and axial charges in the continuum limit. 
While each of the two charges generates an ordinary U(1) symmetry satisfying all the locality properties, they do \textit{not} commute on a finite-size lattice. 
Together they form a non-abelian algebra, known as the Onsager algebra \cite{PhysRev.65.117}, which is consistent with the Nielsen-Ninomiya theorem and matches the continuum anomaly (see Figure \ref{fig:QVQA}).

\section{Symmetries and anomalies in the continuum}
We start with a review of the symmetries and anomalies of the massless Dirac fermion field theory in 1+1D. 
We denote the left- and right-moving (complex, one-component) Weyl fermions as ${\Psi_\text{L}(t,x)}$ and ${\Psi_\text{R}(t,x)}$, respectively. In Minkowski spacetime with metric ${\eta_{\mu\nu} = (1,-1)}$, 
the action for a free, massless Dirac fermion ${\Psi =(\Psi_\text{R} ~\Psi_\text{L})^T}$ is
\ie\label{eq:action}
S&=i \int \mathrm{d}t\,\mathrm{d}x \, \bar\Psi \, \Gamma^\mu\partial_\mu \Psi\\
&=
i \int \mathrm{d}t\,\mathrm{d}x \left[
\Psi_\text{L}^\dagger (\partial_t-\partial_x)\Psi_\text{L}^{\,}
+\Psi_\text{R}^\dagger (\partial_t+\partial_x)\Psi_\text{R}^{\,}
\right] \,,
\fe
where our conventions for the Dirac matrices are ${\Gamma^0=\sigma^x}$, ${\Gamma^1 = -i\sigma^y}$ and ${\bar\Psi =\Psi^\dagger \Gamma^0}$. 
The internal global symmetry for the right-moving Weyl fermion is ${\text{O(2)}^\text{R}=\rm{U(1)}^\text{R}\rtimes \mathbb{Z}_2^{\mathcal{C}^\text{R}}}$, where ${\rm{U(1)}^\text{R}}$ and the charge conjugation symmetry $\mathbb{Z}_2^{\mathcal{C}^\text{R}}$ act on the fermion as ${e^{i\varphi \mathcal{Q}^\text{R}}\Psi_\text{R}e^{-i\varphi \mathcal{Q}^\text{R}}= e^{-i \varphi } \Psi_\text{R}}$ and ${\mathcal{C}^\text{R}\Psi_\text{R}(\mathcal{C}^\text{R})^{-1}= \Psi_\text{R}^\dagger}$, respectively. 
Here $\mathcal{Q}^\text{R}$ is the quantized charge for the right movers, which obeys ${\mathcal{C}^\text{R} \mathcal{Q}^\text{R} (\mathcal{C}^\text{R})^{-1} = -\mathcal{Q}^\text{R}}$. 
A similar global $\rm{O(2)}^\text{L}$ symmetry applies to the left movers, making the total internal global symmetry ${\rm{O(2)}^\text{L} \times \rm{O(2)}^\text{R}}$.

The (quantized) vector and axial charges are defined as ${\mathcal{Q}^\text{V} = \mathcal{Q}^\text{L}+\mathcal{Q}^\text{R}}$ and ${\mathcal{Q}^\text{A} = \mathcal{Q}^\text{L}-\mathcal{Q}^\text{R}}$ and act on the fermions as 
\ie\label{eq:QVA}
&[\mathcal{Q}^\text{V}, \Psi_\text{L}^\dagger] = \Psi^\dagger_\text{L}\,,\qquad
[\mathcal{Q}^\text{V}, \Psi_\text{R}^\dagger] = \Psi_\text{R}^\dagger\,,\\
&[\mathcal{Q}^\text{A}, \Psi_\text{L}^\dagger] = \Psi_\text{L}^\dagger\,,\qquad
[\mathcal{Q}^\text{A}, \Psi_\text{R}^\dagger] =- \Psi^\dagger_\text{R}\,.
\fe
The $\mathbb{Z}_2$ subgroups of the vector and axial U(1) symmetries act identically on the fermions as a fermion parity, so the global form of these symmetries is ${[\text{U(1)}^\mathcal{V} \times \text{U(1)}^\mathcal{A}]/\mathbb{Z}_2}$. 
The vector and axial charges are related by:
\ie\label{eq:CR}
\mathcal{Q}^\text{A} = \mathcal{C}^\text{R} \mathcal{Q}^\text{V}(\mathcal{C}^\text{R})^{-1}\,.
\fe

The vector and axial U(1) symmetries are separately anomaly-free in the sense that there is no obstruction to gauging either one of the two. 
However, there is a mixed anomaly between them, which implies that when the vector symmetry is gauged ({\it i.e.}, in QED), the axial symmetry is broken, and vice versa \cite{PhysRev.128.2425,JOHNSON1963253}. 
This can be seen from the anomalous conservation equation of the axial current $j^{\text{A}}_{\mu}=\bar\Psi \,\Gamma^5 \Gamma_\mu  \Psi$ (with $\Gamma^5=\Gamma^0 \, \Gamma^1$):
\ie\label{eq:anomaly}
\partial^\mu j^{\text{A}}_\mu = -{\frac 1\pi }E\,,
\fe
where $E$ is the electric field. 

\section{Axial charges of staggered fermions}

Consider the 1+1D staggered fermion Hamiltonian~\cite{Kogut:1974ag,Banks:1975gq,Susskind:1976jm} 
\ie \label{eq:Dirac}
H = -i \sum_{j=1}^{L}
\left(
c_j^\dagger c_{j+1}^{\,} + c_{j}^{\,} c_{j+1}^\dagger
\right)  ,
\fe 
where there is a single-component complex fermion $c_j$ at every site $j$, satisfying ${\{c_j, c_{j'}\} = \{c_j^\dagger, c_{j'}^\dagger\} = 0 }$ and ${\{c_j, c_{j'}^\dagger\} = \delta_{j,j'}}$. 
This Hamiltonian is ultra-local, with only nearest-neighbor couplings.\footnote{A Hamiltonian is called ``ultra-local" if every term only involves operators in a finite neighborhood whose size does not scale with the overall system size. In some literature, this property is simply referred to as ``local", while in others a ``local" Hamiltonian can have long-range interaction which decays exponentially in the distance.}
The continuum limit is a single, free, massless Dirac fermion~\eqref{eq:action}. 
We consider both periodic and anti-periodic boundary conditions on a closed chain with $L$ sites, {\it i.e.}, 
${c_{j+L} = (-1)^\nu c_j }$ with ${\nu=0,1}$. 
We assume $L$ to be even for simplicity.
The Hilbert space is  $2^L$ dimensional.

It will be important for the following discussion to decompose the complex fermion into two real fermions as 
${c_j= \frac{1}{2}\left(a_j +i b_j\right)}$, 
where ${a_j^{\,}=a_j^\dagger}$ and ${b_j^{\,}=b_j^\dagger}$ are decoupled Majorana fermions satisfying $\{a_j,a_{j'}\} = \{b_j,b_{j'}\} = 2 \del_{j,j'}$. 
In terms of these Majorana fermions, the Hamiltonian~\eqref{eq:Dirac} becomes
\ie \label{eq:Maj}
H = -\frac i2\sum_{j=1}^{L}(a_ja_{j+1}+ b_jb_{j+1}).
\fe

The Hamiltonian~\eqref{eq:Dirac} has a manifest U(1)$^\text{V}$ fermion-number symmetry whose quantized charge is\footnote{We choose the constant in $\QV$ so that $C \QV C^{-1} =-\QV$. Here, $C$ is charge conjugation, which acts on the fermion as $C c_j C^{-1} =c_j^\dagger$. It becomes the non-chiral charge conjugation $\mathcal{C}^\text{L}\mathcal{C}^\text{R}$ in the continuum limit. 
The axial charge $\QA$ defined later in \eqref{eq:QAdef} also obeys $C \QA C^{-1} =-\QA$.}
\ie \label{eq:QVdef}
\QV = \sum_{j=1}^L \left( c_j^\dagger c_j^{\,} -\frac12 \right) = \frac i2 \sum_{j=1}^L a_j b_j 
\equiv \sum_{j=1}^L \qV_j.
\fe 
It flows to the vector charge $\mathcal{Q}^\text{V}$ of the Dirac fermion field theory in the continuum limit.\footnote{Throughout this paper, we work in units where the lattice spacing is set to be 1. In these units, the continuum limit is achieved by sending $L\to \infty$ while restricting to the low-energy spectrum.} We will hence refer to it as the vector charge or the fermion number. 
The U(1)$^\text{V}$ symmetry acts on the lattice fermions as ${e^{i\varphi \QV} c_j e^{-i\varphi \QV} = e^{-i\varphi}\ c_j}$, 
or equivalently, as
\ie 
e^{i\varphi \QV} a_j e^{-i\varphi \QV} &= \cos\varphi\ a_j + \sin\varphi\ b_j \, ,\\
e^{i\varphi \QV} b_j e^{-i\varphi \QV} &= \cos\varphi\ b_j - \sin\varphi\ a_j \, .
\fe

\subsection{Quantized axial charge}

Is the axial symmetry exact on the lattice? We claim that the lattice operator
\ie \label{eq:QAdef}
\QA
&= \frac12\sum_{j=1}^{L} \left(c_j^{\,}+c_j^\dagger\right)\left(c_{j+1}^{\,}-c_{j+1}^\dagger\right) 
\\
&= \frac i2 \sum_{j=1}^{L} a_jb_{j+1} 
\equiv \sum_{j=1}^L \qA_{j+\frac12}
, 
\fe
obeys the following properties:
 (a) It is quantized with integer eigenvalues. (b) It commutes with the ultra-local Hamiltonian~\eqref{eq:Dirac}. (c) It is a sum of local charge density operators $q^\text{A}_{j+\frac12}$.
 (d) It is bilinear in the fermions. (e) It becomes the continuum axial charge $\mathcal{Q}^\text{A}$ in the continuum limit. 
Property (a)
follows from the fact that the local factors $q^\text{A}_{j+\frac12}$, which are the lattice version of the (time component of the) Noether current, commute with each other and can be simultaneously diagonalized. Properties (b), (c), and (d) are straightforward to verify. Below, we will argue for the last property.

To find a lattice axial charge, we note that 
since the vector charge is manifest on the lattice, using~\eqref{eq:CR}, it suffices to find a lattice realization of the right-moving charge conjugation $\mathcal{C}^\text{R}$. 
The latter is a chiral fermion parity that flips the sign of a single Majorana-Weyl fermion. 
Specifically, we decompose the Weyl fermion into two Majorana-Weyl fermions as ${\Psi_\text{R}=\lambda_\text{R} + i \chi_\text{R}}$, then $\mathcal{C}^\text{R}$ only flips the sign of $\chi_\text{R}$. 
It is known that such a chiral fermion parity is realized as a lattice translation on the lattice, which we review below. 

We focus on the Majorana fermion $b_j$ in the lattice Hamiltonian~\eqref{eq:Maj}, which flows to the left and right Majorana-Weyl fermions $\chi_\text{L}$ and $\chi_\text{R}$ in the continuum. 
Up to a constant, the Hamiltonian for $b_j$ in momentum space is 
$ {\sum_{0< k <L/2} \sin  \frac{2\pi k}L \beta_{-k} \beta_{k}}$, where   
${\beta_k  = \frac{1}{\sqrt{L}}\sum_{j=1}^L e^{-\frac{2\pi i k j}L} b_j}$ 
with  ${k\in \mathbb{Z}+\nu/2}$ and $k\sim k+L$.
The ground state(s) $\ket{\Omega}$ obeys ${\beta_{k}\ket{\Omega} =0}$ for all ${0< k < \frac L2}$. The left- and right-moving modes in the continuum field theory created by $\chi_\text{L}$ and $\chi_\text{R}$ arise from the lattice modes created by $\beta_{-k}$ with $k$ close to ${k=0}$ and ${k=\frac L2}$, respectively.

The key symmetry in the lattice model~\eqref{eq:Maj} is the translation operator $T_b$ that shifts the $b_j$ fermion by one site but leaves $a_j$  invariant: 
\ie
T_b^{\,} a_j T_b^{-1}=a_j\,,\qquad T_b^{\,} b_j T_b^{-1}=b_{j+1}.
\fe
In momentum space, 
$T_b \beta_k T_b^{-1} = e^{\frac{2\pi i k}L} \beta_k$, hence $T_b$ acts with a relative minus  sign between the modes around ${k=0}$ and ${k=\frac L2}$ in the  limit $L\to \infty$. 
We conclude that on the low-lying states,  $T_b$ acts as the right-moving charge conjugation times the continuum translation operator~\cite{Seiberg:2023cdc}:
\ie\label{eq:emanant}
T_b = \mathcal{C}^\text{R} e^{\frac{2\pi i \mathcal{P}}L}
\fe
where $\mathcal{P}$ is the continuum momentum operator acting only on the Majorana fermion $\chi$ (but not $\lambda$).\footnote{This is to be contrasted with the lattice translation  $T=T_aT_b$ that shifts both $a_j$ and $b_j$ by one site. In the continuum limit, $T$  flips the signs of both $\lambda_\text{R},\chi_\text{R}$~\cite{Banks:1975gq,Susskind:1976jm}, while $T_b$ only flips the sign of $\chi_\text{R}$. See Refs.~\cite{Seiberg:2023cdc,Clancy:2023ino} for recent discussions of chiral fermion parities from the lattice.}

Having identified the lattice origin of the right-moving charge conjugation $\mathcal{C}^\text{R}$, we follow~\eqref{eq:CR} to define the lattice axial charge:
\ie \label{eq:QVQA}
\QA  = T_b^{\,} \QV T_b^{-1}\, ,
\fe
which gives~\eqref{eq:QAdef}. 
On the low-lying states in the $L\to \infty $ limit,  $e^{2\pi i \mathcal{P}/L}$ becomes 1 and $T_b \sim \mathcal{C}^\text{R}$, so from~\eqref{eq:CR} we see that $\QA$ becomes the continuum axial charge $\mathcal{Q}^\text{A}$. 
From this expression, it is clear that $\QA$ has integer eigenvalues since it is unitarily equivalent to $\QV$. It is also clear that it commutes with the Hamiltonian since $\QV$ and $T_b$ do.

Since $\QA$ is quantized, we can exponentiate it to find an exact   U(1)$^\text{A}$ axial symmetry on the lattice, which acts locally on the fermions as:
\ie 
e^{i\varphi \QA} a_j e^{-i\varphi \QA} &= \cos\varphi\ a_j + \sin\varphi\ b_{j+1} \, ,\\
e^{i\varphi \QA} b_j e^{-i\varphi \QA} &= \cos\varphi\ b_j - \sin\varphi\ a_{j-1} \, ,
\fe
or equivalently, as
\begin{multline}
e^{i\varphi \QA} c_j e^{-i\varphi \QA} = \cos\varphi \ c_j \\ 
- i \sin\varphi \ \frac12\left(c_{j-1}^{\dagger} + c_{j-1}^{\,} - c_{j+1}^{\dagger} + c_{j+1}^{\,} \right) \, .
\end{multline} 
This quantized axial charge was first identified in \cite{Thacker:1994ns,Horvath:1998gq} from the connection to integrability. 
Here, we provide an alternative derivation using lattice translation.

While each of $\QV$ and $\QA$ generates an ordinary U(1) global symmetry, interestingly, these two lattice charges do \textit{not} commute:  
\ie \label{eq:QVQAcom}
[\QV,\QA] = -\sum_{j=1}^L \left(c_j^{\,} c_{j+1}^{\,} 
+ c_j^\dagger 
c_{j+1}^\dagger
\right) \, .
\fe 
This is to be contrasted with the continuum where ${[\mathcal{Q}^\text{V},\mathcal {Q}^\text{A}]=0}$.  
In Appendix~\ref{app:commutator}, we show that the non-vanishing lattice commutator goes to zero on the low-lying states in the ${L\to \infty}$ limit. 
Note that ${(-1)^F\equiv e^{i\pi \QV}=e^{i\pi \QA}}$ is the (non-anomalous) fermion parity that flips the sign of all the fermions. It  commutes with both U(1)'s, {\it i.e.}, 
${[(-1)^F,\QA] = [(-1)^F,\QV] = 0}$.

\subsection{Unquantized axial charge}

Although ${[\QA,\QV]\neq 0}$, there is another conserved operator ${\tQA =\frac12 \big( \QA + e^{\frac{i\pi}2 \QV} \QA e^{-\frac{i\pi}2 \QV} \big) }$ that obeys ${[\tQA , \QV]=0}$.  
This operator, which is the most straightforward lattice regularization of the axial charge~\cite{Banks:1975gq,Susskind:1976jm}, can also be written as a sum of local terms: 
\ie \label{eq:tQAdef}
\tQA &=
\frac12 \sum_{j=1}^{L} \left(
c_{j}^\dagger c_{j+1}^{\,} -
c_{j}^{\,}c_{j+1}^\dagger  
\right)
\\
&= \frac i4 \sum_{j=1}^{L} \left(a_jb_{j+1} - b_ja_{j+1}  \right)
\equiv \sum_{j=1}^L \tqA_{j+\frac12}.
\fe 
In the continuum limit, $\QV$ and $\QA$ commute, and $\tQA$ flows to the same continuum axial charge $\mathcal{Q}^\text{A}$. 
However, the price we pay is that the eigenvalues of $\tQA$ are generally irrational and are not quantized.\footnote{Up to an overall constant,  $\tQA$ is equivalent to the Hamiltonian~\eqref{eq:Dirac} for ${L=0}$ mod 4, under the unitary transformation ${c_j\mapsto i^jc_j}$.} Furthermore, while ${e^{i\lm\tQA}}$ is a conserved unitary operator with ${\lm\in\mathbb{R}}$, it fails to send local operators to local operators when ${\lm\sim\mathcal{O}(L)}$ (i.e., it is not a locality-preserving unitary). See Appendix~\ref{actionApp} for further discussion.

The commutator between the local charge densities $\qV_j$ and $\tqA_{j'+\frac12}$ is~\cite{SS}\footnote{We thank Tom Banks and Nathan Seiberg for discussions on the Schwinger term and collaboration on another project.}
\ie 
[\qV_j,\tqA_{j'+\frac12}]  
&= \frac i2( \del_{j,j'}-\del_{j-1,j'})h_{j'+\frac12} \, ,
\fe 
where ${h_{j+\frac12} = -\frac i2( a_j a_{j+1} + b_jb_{j+1})}$ is the Hamiltonian density. This is the lattice avatar of the Schwinger term, which encodes the mixed anomaly of U(1)$^\mathcal{V}$ and U(1)$^\mathcal{A}$ in the continuum.

The simultaneous realization of both vector and axial U(1) symmetries seems to conflict with the well-known Nielsen-Ninomiya theorem. 
However, neither of the conserved lattice operators $\QA$ and $\tQA$ satisfies both conditions (i) and (ii) mentioned in the Introduction: $\QA$ is quantized but does not commute with $\QV$, while $\tQA$ commutes with $\QV$ but is not quantized. 
(See Table \ref{tab:charges}.) 
We, therefore, cannot define a notion of lattice chirality using them. 
The fact that these two conditions cannot be met simultaneously is in harmony with the Nielsen-Ninomiya theorem. 
Relatedly, we do not have U(1)$^\text{L}$ or U(1)$^\text{R}$ symmetries on the lattice (which would have violated the no-go theorem in Ref.~\cite{Kapustin:2024rrm}) since $\frac12(\QV\pm \QA)$ do not have quantized eigenvalues.

\setlength{\tabcolsep}{8pt} \renewcommand{\arraystretch}{1.5} \begin{table}
\begin{tabular}{|l||c|c|} 
\hline  
  lattice operators  &  $\QA$ & $\tQA$ \\ 
\hhline{|=#=|=|}
quantized eigenvalues?  & \checkmark &   \\
\hline 
 $[~ \bullet ~, \QV]=0$? &  & \checkmark \\
\hline 
 $[~ \bullet ~, H]=0$? & \checkmark & \checkmark \\
\hline
sum of local terms?  & \checkmark & \checkmark \\
\hline
\end{tabular}
\caption{\label{tab:charges} Properties of the two lattice operators, both flowing to the same axial charge of a free Dirac fermion in the continuum limit.} 
\end{table} \renewcommand{\arraystretch}{1}

\subsection{From the lattice  to the continuum}
We now present a more detailed analysis in momentum space to relate the lattice and continuum symmetries. 
Let us define ${\gamma_k  = \frac{1}{\sqrt{L}}\sum_{j=1}^L e^{-\frac{2\pi i k j}L} c_j}$, which satisfies 
${\{\gamma_k,\gamma_{k'}\}=\{\gamma_k^\dagger , \gamma^\dagger_{k'}\}=0}$ and ${\{\gamma^{\,}_k ,\gamma^\dagger_{k'}\}=\delta_{k,k'}}$.\footnote{They are related to the Fourier modes $\al_k,\beta_k$ of $a_j,b_j$ by ${\gamma_k^{\,} =\frac12 (\al_k^{\,} + i \bt_k^{\,}) , 
\gamma_k^\dagger =\frac12 (\al_{-k}^{\,} - i \bt_{-k}^{\,})}$. }
The Hamiltonian~\eqref{eq:Dirac} in momentum space is ${H = \sum_{-\frac L2<  k\leq \frac L2} 2 \sin\frac{2\pi k}{L}\  \gamma_k^\dagger \gamma_k}$. 
The ground state(s)   is(are) annihilated by
$\gamma_k^{\,} $ with ${0<k<\frac L2}$,  and by 
$\gamma_k^{\dagger}$ with  ${-\frac L2<k<0}$.
For periodic boundary condition ${\nu=0}$, there are four degenerate ground states due to the zero modes at ${k=0}$ and $\frac{L}{2}$. 
For anti-periodic boundary condition  ${\nu=1}$, there is a unique ground state.

The conserved operators in momentum space are
\ie
&{\QV = \sum_{k} \left( \gamma_k^\dagger \gamma_k^{\,} -\frac12 \right)},\\ 
&{\tQA = \sum_{k} \cos\frac{2\pi k}{L}\ \gamma_k^\dagger \gamma_k^{\,}}\\
&{\QA = \tQA + 
\frac12 \sum_{k}\left(e^{-\frac{2\pi i}{L}k}\ \gamma_k \gamma_{-k} 
+ \hc \right)}
\fe
where all the sums are over ${ -\frac L2<  k\leq \frac L2}$.
The middle equation shows that the eigenvalues of $\tQA$ are not quantized, as mentioned earlier.

In the continuum limit, we only retain the Fourier modes $\gamma_k^\dagger$, $\gamma_{-k}^{\,}$, $\gamma_{\frac L2 -k}^\dagger$, and $\gamma_{-\frac L2 +k}^{\,}$ with ${|k|\ll L}$, which become the creation operators for $\Psi^\dagger_{\text{L}}$, $\Psi^{\,}_{\text{L}}$, $\Psi^\dagger_{\text{R}}$, and $\Psi^{\,}_{\text{R}}$, respectively. 
We now compute the commutators of the unquantized $\tQA$ and quantized $\QA$ axial charges with the creation operators:
\ie 
&[\tQA, \gamma^\dagger_k] = \cos \frac{2\pi k}{L} \gamma^\dagger_k  \,,\\
&[\QA, \gamma^\dagger_k] = \cos \frac{2\pi k}{L} \gamma^\dagger_k + i \sin \frac{2\pi k}{L} \gamma^{\,}_{-k} \,,
\fe 
while those with $\gamma_k$ can be obtained similarly. 
Hence, for finite $k$ and $L\to\infty$, 
\ie
&\lim_{L\to \infty}[\QA, \gamma^\dagger_k] =\lim_{L\to \infty} [\tQA,\gamma^\dagger_k] =\gamma^\dagger_k,\\ 
&\lim_{L\to \infty}[\QA, \gamma^\dagger_{\frac L2-k}] =\lim_{L\to \infty}[\tQA,\gamma^\dagger_{\frac L2-k}] = -\gamma^\dagger_{\frac L2-k}.
\fe
Comparing with~\eqref{eq:QVA}, this proves that both $\QA$ and $\tQA$ flow to the continuum axial charge $\mathcal{Q}^\text{A}$ in the limit. 
As a demonstration, we compute the eigenvalues and expectation values of these conserved charges on the low-lying lattice states in Table \ref{tab:lowstates}.

\setlength{\tabcolsep}{8pt} \renewcommand{\arraystretch}{1.5} \begin{table}[t]
\begin{tabular}{|c|c|c|c||c|} 
\hline  
lattice   & $\QV$ & $\tQA$ & $\langle \QA\rangle$ & cont.  \\ 
states&&&&op.\\
\hhline{|=|=|=|=#=|}
$\ket{\Omega}$  & $0$ & $0$& $0$& 1 \\
\hline
$\gamma_{\frac12}^\dagger\ket{\Omega}$  & $+1$ & $\cos \frac\pi L $& $\cos \frac\pi L$ & $\Psi_\text{L}^\dagger$\\
\hline
$\gamma_{-\frac12}\ket{\Omega}$ & $-1$ & $-\cos \frac\pi L $& $-\cos \frac\pi L$ & $\Psi_\text{L}$ \\
\hline
$\gamma_{\frac L2-\frac12}^\dagger\ket{\Omega}$ & $+1$ & $-\cos \frac\pi L $& $-\cos \frac\pi L$  & $\Psi_\text{R}^\dagger$\\
\hline
$\gamma_{-\frac L2+\frac12}\ket{\Omega}$& $-1$ & $\cos \frac\pi L $& $\cos \frac\pi L$  & $\Psi_\text{R}$ \\
\hline
\end{tabular}
\caption{\label{tab:lowstates} The eigenvalues of the vector charge $\QV$ and the unquantized axial charge $\tQA$ for the low-lying energy eigenstates of the Hamiltonian~\eqref{eq:Dirac} with anti-periodic boundary condition. 
Since the quantized axial charge $\QA$ does not commute with $\QV$ or $\tQA$, it cannot be simultaneously diagonalized; instead, we record its expectation values. The rightmost column shows the local operators that create these states in the continuum limit. } 
\end{table} \renewcommand{\arraystretch}{1}

\subsection{Chiral anomaly from the Onsager algebra}

We now discuss the non-abelian algebra generated by the charges $\QV$ and $\QA$. 
Let us define
$G_n = \frac{i}2\sum_j( a_j a_{j+n} - b_j b_{j+n})$ and 
$ Q_n =
\frac{i}{2} \sum_j a_j b_{j+n}$ with ${n\in \mathbb{Z}}$. 
These operators all commute with the Hamiltonian and obey the following closed algebra:
\ie \label{eq:nonabelian}
[Q_n,Q_m] &= iG_{m-n}
\, ,~~~[G_n,G_m]=0\,, \\ 
[Q_n,G_m] & =2 i(Q_{n-m}-Q_{n+m}) \, ,
\fe 
which is precisely the  Onsager algebra \cite{PhysRev.65.117}.\footnote{We thank Yuan Miao for pointing this out to us.} 
In particular, ${\QV=Q_0}$, ${ \QA=Q_1}$, ${\tQA=\frac12 (Q_1+Q_{-1})}$.
While ${Q_n = T_b^n Q_0 T_b^{-n}}$ have integer eigenvalues, $G_n$ do not.

In the continuum limit, the operators in the Onsager algebra with a fixed finite $n$ become
\ie
\lim_{L\to \infty} G_n =0,~~~
\lim_{L\to \infty} Q_n = \begin{cases}
\mathcal{Q}^\text{V}\qquad&\text{for $n$ even},\\
\mathcal{Q}^\text{A}\qquad&\text{for $n$ odd},
    \end{cases}
\fe
where  we compute the limit of $G_n$ in Appendix \ref{app:commutator}, and the limit of $Q_n$ follows from ~\eqref{eq:emanant}.
Therefore, we find an infinite tower of conserved lattice operators, with $\QA$ and $\tQA$ special cases of them, that flow to the same axial charge in the continuum limit. 
The anomalous ${[\text{U(1)}^\mathcal{V} \times \text{U(1)}^\mathcal{A}]/\mathbb{Z}_2}$ symmetry in the massless Dirac fermion field theory arises from the Onsager algebra in the staggered fermion lattice model.

Since the model~\eqref{eq:Dirac} is integrable,  it is expected to have many conserved quantities. 
However, the majority of these conserved quantities are non-local, meaning they cannot be expressed as a sum of operators supported in a finite region, nor do they map local operators to other local operators. The interesting point, however, is that among these conserved 
 operators in the Onsager algebra, ${\QA}$ (along with ${Q_n}$ for small, odd ${n}$) is local and flows to the axial charge in the continuum, which has an anomaly with the vector charge.

\subsection{Comparison with other constructions}

Many of the discussions of lattice chiral symmetries are in the context of Euclidean lattice models, while ours concerns the Hamiltonian lattice models. We will not attempt to give a comprehensive comparison with all the existing constructions, but refer the readers to Ref.~\cite{Kaplan:2009yg} for a review. 

The Euclidean version of the staggered fermion, which is reviewed in Ref.~\cite{Golterman:2024xos}, has two U(1) symmetries. The first is the manifest vector, or fermion number, symmetry. 
The second is sometimes referred to as the U(1)$_\epsilon$ symmetry \cite{Karsten:1980wd,Golterman:1984cy}. 
In contrast to the symmetries in the Hamiltonian model, these two U(1) symmetries commute in the Euclidean model. 
However, U(1)$_\epsilon$ acts with an alternating sign in the Euclidean time direction, and therefore does not lead to an ordinary global symmetry in the Hamiltonian formalism.
It would be interesting to further explore the connection between U(1)$_\epsilon$ and our non-abelian algebra, especially along the lines of Ref.~\cite{Catterall:2022jky}.

Next, we compare this discussion with other Hamiltonian lattice models in the literature.  
Refs.~\cite{Drell:1976mj,Karsten:1978nb,Weinstein:1982ht} discuss an axial charge that is both quantized and commutes with the vector charge, but their Hamiltonian is not local, and the  Nielsen-Ninomiya theorem does not apply. 
The quantized axial charge was pointed out in Refs.~\cite{Thacker:1994ns,Horvath:1998gq}, and was referred to as the ``arrow charge" in the latter reference in connection with the integrability literature.
In this letter, we provided an alternative derivation of this quantized axial charge, from which we found a family of other lattice realizations and identified the non-abelian Onsager algebra they generate. 
The unquantized axial charge was first discussed in Refs.~\cite{Banks:1975gq,Susskind:1976jm}, and was later elaborated in   Ref.~\cite{Horvath:1998gq,Creutz:2001wp}  in the context of the Hamiltonian version of the overlap formulation~\cite{Narayanan:1994gw,Neuberger:1997fp,Neuberger:1998wv} and the Ginsparg-Wilson relation~\cite{PhysRevD.25.2649}.
Finally, the Hamiltonian in  \eqref{eq:Dirac} is in 1+1D with nearest-neighbor couplings, contrasting with the domain wall fermion approach where the model is effectively 2+1D~\cite{Kaplan:1992bt}. (See also Refs.~\cite{Kaplan:2023pvd,Kaplan:2023pxd} for recent progress.)

\section{Coupling to gauge fields}

\subsection{Gauging the vector symmetry}

Let us now discuss the fate of the two axial charges after we gauge U(1)$^\text{V}$. The gauged Hamiltonian is a lattice regularization of 1+1D QED, {\it i.e.}, the Schwinger model~\cite{PhysRev.128.2425}. 
We introduce a U(1)-valued gauge field ${U_{j+\frac 12}}$  and an integer-valued electric field operator ${L_{j+\frac 12}}$ on each link. They satisfy $[L_{j+\frac12},U_{j'+\frac12}] = \del_{j,j'} U_{j'+\frac12}$. The gauged Hamiltonian is~\cite{Banks:1975gq,Carroll:1975gb}
\ie \label{eq:HV}
H_\text{V} =  &-i \sum_j \left(c_j^\dagger U_{j+\frac 12}^{\,} c_{j+1}^{\,} + c_j^{\,} U_{j+\frac 12}^\dagger c_{j+1}^\dagger \right)
\\ 
&
+K\sum_j L_{j+\frac 12}^2 \,.
\fe

Furthermore, we impose the Gauss constraint, $L_{j+\frac12}-L_{j-\frac12} = q^\text{V}_{j+\frac12} + \frac{(-1)^j}{2}$~\cite{Hamer:1997dx,Dempsey:2022nys}. 
The Gauss law restricts the Hilbert space to a subsector of fixed $\QV$ charge. Since the quantized axial charge $\QA$ does not commute with $\QV$, it does not act within the gauge-invariant subspace. 
$\QA$ is therefore no longer a symmetry when we couple to the dynamical gauge field for U(1)$^\text{V}$. This is analogous to gauging a U(1) subgroup of SU(2): if the $S^z$ symmetry is gauged, the $S^x$ and $S^y$ charges cannot be made gauge-invariant and are, therefore, explicitly broken.

On the other hand, the unquantized axial charge $\tQA$ can be made gauge-invariant as $\tQA(U) = \frac12 \sum_j (\,
c_j^\dagger U_{j+\frac 12}^{\,} c_{j+1}^{\,} - c_j^{\,} U_{j+\frac 12}^\dagger c_{j+1}^\dagger  \,)$.\footnote{The continuum counterpart is the following: in the point-splitting regularization of the axial current  $j^\text{A}$, an infinitesimal Wilson line must be introduced to ensure gauge invariance~\cite{SHIFMAN1991341}.} 
Even though it is a gauge-invariant operator acting in the gauged theory, it fails to commute with the gauged Hamiltonian, 
\ie 
-i\, [\tQA(U),H_\text{V}] = -\frac {K}2 \sum_j \{L_{j+\frac12}, h_{j+\frac12}(U) \} \, ,
\fe 
where ${h_{j+\frac12}(U) =-i\left(c_j^\dagger U_{j+\frac 12}^{\,} c_{j+1}^{\,} + c_j^{\,} U_{j+\frac 12}^\dagger c_{j+1}^\dagger \right)}$ is the Hamiltonian density for the fermions.
This is the lattice avatar of the Schwinger anomaly of the continuum axial charge ${\frac{d}{dt}\cQA =-i [\cQA,H] = -{\frac 1\pi}\int \mathrm{d}x\, E}$~\cite{Manton:1985jm}, which is the integrated form of~\eqref{eq:anomaly}. 
We conclude that both the quantized $\QA$ and the unquantized $\tQA$ axial symmetries are broken as we gauge the vector U(1)$^\text{V}$ symmetry.

\subsection{Gauging the axial symmetry}

The main advantage of the quantized axial charge $\QA$, compared to the unquantized one $\tQA$, is that it can be coupled to dynamical (compact) U(1) gauge fields. 
The gauged Hamiltonian is  obtained by conjugating   $H_\text{V}$ with $T_b$:
\ie\label{eq:HA}
H_\text{A} = &\sum_j 
\Big[
- \frac i4
(a_j -ib_{j+1} )U_{j+\frac12} (a_{j+1} +i b_{j+2})
\\
&\qquad +\hc \Big]
+K \sum_j L_{j+\frac12}^2\,.
\fe
The Gauss law is obtained similarly, and found to be ${L_{j+\frac12}-L_{j-\frac12} = \qA_{j+\frac12} + \frac{(-1)^j}{2}}$. 
We conclude that $\QV$ and $\QA$ can individually be gauged on the lattice and are free of self-anomalies.

\section{A lattice anomaly as an obstruction to gapped phases}

Do the lattice axial and vector symmetries have an anomaly? 
Conventionally, the anomaly of a global symmetry is defined as the obstruction to gauging the symmetry. 
However, the lattice charges $\QA$ and $\QV$ generate the non-abelian Onsager algebra~\eqref{eq:nonabelian}, which includes highly non-local charges, so it is not clear if there is a sensible prescription for gauging it.  
On the other hand, it has been advocated in Refs.~\cite{Chang:2018iay, Wen:2018zux, Thorngren:2019iar, Choi:2021kmx} that a global symmetry should be called anomalous if there does not exist a trivially gapped phase (\textit{i.e.}, a gapped phase with a non-degenerate ground state and no long-range entanglement) preserving the symmetry. 
This definition is inspired by the 't Hooft anomaly matching argument and avoids the need to discuss the gauging of said global symmetry. 
Below, we will show that the lattice axial and vector symmetries together are anomalous in this sense.
See Refs.~\cite{Komargodski:2020mxz, Choi:2023xjw, Choi:2023vgk} for the relation between these two definitions of anomalies. 

Concretely, we prove in Appendix~\ref{app:symdef} that local deformations of the Hamiltonian~\eqref{eq:Dirac} preserving both $\QV$ and $\QA$ are necessarily quadratic 
in the fermions.
In fact, the only $\QV$ and $\QA$ symmetric local deformations are of the form 
(${n\geq 2}$)
\ie 
\del H_n &= 
\sum_{j=1}^L
-i(c_j^\dagger c_{j+n}^{\,} + c_j^{\,} c_{j+n}^\dagger ) \,,
\fe 
which flow to irrelevant deformations of the Dirac conformal field theory (CFT) in the IR. 
For a small deformation strength, such terms only renormalize the velocities of the left- and right-moving fermions in the continuum limit.

In particular, the quartic Thirring coupling ${c_j^\dagger c_j^{\,} c_{j+1}^\dagger c_{j+1}}$ preserves $\QV$ but breaks $\QA$. 
This is to be contrasted with its continuum limit; the continuum Thirring coupling  $\frac14(\bar\Psi \Gamma^\mu \Psi)(\bar\Psi \Gamma_\mu \Psi) 
=\Psi^\dagger_L \Psi^{\,}_L \Psi^\dagger_R \Psi^{\,}_R$ preserves both the vector and axial symmetries $\mathcal{Q}^\text{V},\mathcal{Q}^\text{A}$, and leads to an exactly marginal deformation of the Dirac fermion CFT.

One interesting corollary is that \textit{any Hamiltonian that commutes with both $\QV$ and $\QA$ must be gapless}. 
This is reminiscent of the constraint from perturbative anomalies in continuum field theory. 
In continuum field theory, perturbative anomalies of continuous global symmetries are encoded in the local operator product expansion and, therefore, cannot be matched by a gapped phase. 
Even when the symmetry is spontaneously broken, there are gapless Goldstone boson modes. 
The symmetries $\QV$ and $\QA$ present an example of such a gapless constraint on a lattice with a finite-dimensional local Hilbert space.

This constraint is in fact much stronger than the typical discrete anomalies or the Lieb-Schultz-Mattis constraints~\cite{LIEB1961407,2000PhRvL..84.3370O,Hastings:2003zx}, where the low-energy phase is constrained to be either gapless, \textit{or} gapped with some nontrivial features, such as degenerate ground states and/or long-range entanglement described by a topological quantum field theory. It is also stronger than the ``symmetry-enforced gaplessness" discussed in Refs.~\cite{2014PhRvB..89s5124W,2016PhRvB..94x5107W,2017PhRvB..95h5135S,Wang:2017txt,Cordova:2019bsd,Cordova:2019jqi,Apte:2022xtu}, where a gapped phase with spontaneous broken discrete symmetries ({\it e.g.}, time-reversal symmetry) remains a possibility. 
The charges $\QV$ and $\QA$ entirely exclude any gapped phases, leaving the gapless phase as the only possibility.

While the preservation of both $\QV$ and $\QA$ imposes a nontrivial gapless constraint, each of these two conserved charges can individually be coupled to respective gauge fields (as in \eqref{eq:HV} and \eqref{eq:HA}) and is free of anomalies. Moreover, it is possible to deform the gapless Hamiltonian~\eqref{eq:Dirac} to a trivially gapped phase with a unique vacuum while preserving either $\QV$ or $\QA$ (but not both). 
If we choose to preserve $\QV$, while violating $\QA$, we can add the deformation ${\del H = \sum_j (-1)^j c_j^\dagger c_j^{\,}}$ to open an energy gap. 
In the continuum, this corresponds to deforming the Lagrangian density by the mass term ${\Psi_{\rm R}^\dagger\Psi_{\rm L}^{\,} + \Psi_{\rm L}^\dagger\Psi_{\rm R}^{\,}}$~\cite{Banks:1975gq,Susskind:1976jm}. See Ref.~\cite{Dempsey:2022nys} for the precise relation between this lattice deformation and the continuum mass term. 
On the other hand, the deformation  $T_b\,  \delta H \, T_b^{-1} = \sum_j (-1)^j \qA_{j+\frac 12} $preserves $\QA$ but violates $\QV$ and drives the system to a trivially gapped phase. In the continuum, it corresponds to  ${\Psi_{\rm R}^{\,}\Psi_{\rm L}^{\,} + \Psi_{\rm L}^\dagger\Psi_{\rm R}^\dagger}$.

Finally, we note that the charge $\QV$ is on-site in the sense that the local terms ${\qV_j}$ involve fermions only on site $j$, and it maps a local operator at site $j$ to another at the same location. 
On the other hand, $\QA$ is not on-site since it smears a local operator at site $j$ to its nearest neighbors. 
The unitary transformation by ${T_b^{-1}}$ renders $\QA$ on-site, but this comes at the cost of $\QV$ no longer being on-site. 
However, there does not seem to be a unitary transformation that makes both $\QV$ and $\QA$ on-site simultaneously, hinting at a mixed anomaly between them. See Refs.~\cite{Wang:2013yta, Wang:2018ugf} for related discussions.

\section{Exact winding symmetry in the XX model}
It is well-known that upon bosonization, the fermionic Hamiltonian~\eqref{eq:Dirac} becomes the XX model, ${H = \sum_{j} X_j X_{j+1} + Y_jY_{j+1} }$~\cite{LIEB1961407}. 
The vector charge $\QV$ becomes ${\frac12 \sum_j Z_j}$, the charge of the manifest U(1) spin rotation symmetry of the XX Hamiltonian. 
In the continuum, this flows to the momentum U(1) symmetry of the ${c=1}$ compact boson CFT at radius ${R=\sqrt 2}$.\footnote{As standard in the string theory literature, we refer to the two U(1) symmetries of the compact boson CFT as the ``momentum" and ``winding" symmetries. In some communities, they are respectively referred to as the ``charge" and ``vortex" symmetries. Here ``momentum" refers to that in the target space. In the language of the Luttinger liquid, these two U(1) symmetries respectively shift the two scalar fields. Our convention for the target space radius $R$ is so that $R=1$ is the self-dual point under T-duality, which enjoys an enhanced $su(2)$ chiral algebra.} 
On the other hand, the axial charge $\QA$ is mapped to a second U(1) symmetry of the XX model, which flows to the winding U(1) symmetry in the continuum; its explicit form on a chain with even $L$ is\footnote{In fact, when ${L = 0~\mathrm{mod}~4}$, the XX model is unitarily equivalent to the Levin-Gu spin chain~\cite{CLV1301, LOZ220403131}, and this lattice winding charge becomes the U(1) charge ${\frac14\sum_j Z_j Z_{j+1}}$ of Ref.~\cite{LG1209}. On the other hand, the Levin-Gu spin chain is known to be equivalent to the XX model coupled to a $\mathbb{Z}_2$ gauge field. Therefore, the XX model is self-dual under gauging, leading to a non-invertible symmetry. The latter is the lattice realization of the symmetry discussed in Refs.~\cite{Ji:2019ugf,Thorngren:2021yso,Choi:2021kmx}.}
\ie
\frac14\sum_{j=1}^{L/2}( X_{2j-1}Y_{2j}  -Y_{2j} X_{2j+1}).
\fe
These two lattice charges first appeared in \cite{PhysRev.65.117}, and were later   discussed in \cite{Vernier:2018han,Miao:2021pyh, PZG230314056} in the context of symmetries of the XX model. 
These two charges are locally equivalent to $\QV$ and $\QA$ in the staggered fermion model, but they are different globally. 
The global aspects of bosonization, the exact winding symmetry, and the relation to a non-invertible symmetry of the XX model will be discussed in Ref.~\cite{Pace:2024oys}.

\section{Conclusion}

While a lattice model with a finite-dimensional Hilbert space cannot host the exact chiral anomaly, the symmetry generators for the axial and vector symmetries -- which in the continuum limit have a mixed anomaly -- exist exactly in the staggered fermion lattice model. 
The quantized charges $\QV,\QA$ \cite{Thacker:1994ns,Horvath:1998gq} resemble their continuum counterparts closely: each one of them generates a (compact) U(1) global symmetry and can be gauged. However, they do not commute with each other on the lattice. 
By contrast, the unquantized axial charge $\tQA$ \cite{Banks:1975gq,Susskind:1976jm,Horvath:1998gq,Creutz:2001wp} generates an $\mathbb{R}$ global symmetry, and it is not clear how to couple it to U(1) gauge fields.

It would be interesting to investigate the fate of these charges for interacting fermions {\it e.g.}, in 1+1D QED with multiple flavors of fermions and in QCD, where gauge interactions make the models non-integrable. 
It would also be natural to analyze chiral global symmetries in other models, such as the 3450 model. 
Another generalization is to explore quantized axial charges in 3+1D staggered fermions, which is more phenomenologically relevant.
One qualitative difference compared to the 1+1D case is that the axial symmetry in 3+1D not only has a mixed anomaly with the vector symmetry but it is also anomalous by itself.

The phenomenon discussed here is qualitatively similar to the emanant symmetry of Ref.~\cite{Cheng:2022sgb} and the non-invertible Kramers-Wannier symmetry in Refs.~\cite{Seiberg:2023cdc,Seiberg:2024gek,Gorantla:2024ocs, PGLA240612962}. 
In all these cases, we start with an internal global symmetry $G_\text{IR}$ in a continuum field theory in the IR and seek a UV symmetry $G_\text{UV}$ in a microscopic lattice realization. While the symmetry operators for $G_\text{IR}$ exist exactly on the lattice (and therefore $G_\text{IR}$ is not emergent), they obey a different algebra $G_\text{UV}$ compared to the IR one. In Refs.~\cite{Metlitski:2017fmd,Seiberg:2023cdc,Seiberg:2024gek,Gorantla:2024ocs, PGLA240612962}, the UV symmetries mix with lattice translations. 
In the current letter, the IR symmetry ${G_\text{IR}=[\text{U(1)}^\mathcal{V} \times \text{U(1)}^\mathcal{A}]/{\mathbb{Z}_2}}$, which has an anomaly, emanates~\cite{Cheng:2022sgb} from the non-abelian Onsager algebra $G_\text{UV}$.

\begin{acknowledgments}

We would like to thank \"Omer Aksoy, Tom Banks, Meng Cheng, Will Detmold, Igor Klebanov, Ho Tat Lam, Michael Levin, Laurens Lootens, Yuan Miao, Nathan Seiberg, Hersh Singh, Senthil Todadri, Frank Verstraete, and Yichen Xu for interesting discussions. SHS is especially grateful to Tom Banks and Nathan Seiberg for numerous enlightening discussions on lattice fermions. 
We would also like to express our special gratitude to Hersh Singh for his valuable insights and guidance toward relevant literature. 
We thank Simon Catterall, Aleksey Cherman, Theo Jacobson, Igor Klebanov, and Hersh Singh for helpful comments on a draft. 
AC is supported by NSF DMR-2022428 and by the Simons Collaboration on Ultra-Quantum Matter, which is a grant from the Simons Foundation (651446, Wen).
SDP is supported by the National Science Foundation Graduate Research Fellowship under Grant No. 2141064. 
SHS is supported in part by NSF grant PHY-2210182.

\end{acknowledgments}

\bibliography{refs}

\begin{thebibliography}{81}%
\makeatletter
\providecommand \@ifxundefined [1]{%
 \@ifx{#1\undefined}
}%
\providecommand \@ifnum [1]{%
 \ifnum #1\expandafter \@firstoftwo
 \else \expandafter \@secondoftwo
 \fi
}%
\providecommand \@ifx [1]{%
 \ifx #1\expandafter \@firstoftwo
 \else \expandafter \@secondoftwo
 \fi
}%
\providecommand \natexlab [1]{#1}%
\providecommand \enquote  [1]{``#1''}%
\providecommand \bibnamefont  [1]{#1}%
\providecommand \bibfnamefont [1]{#1}%
\providecommand \citenamefont [1]{#1}%
\providecommand \href@noop [0]{\@secondoftwo}%
\providecommand \href [0]{\begingroup \@sanitize@url \@href}%
\providecommand \@href[1]{\@@startlink{#1}\@@href}%
\providecommand \@@href[1]{\endgroup#1\@@endlink}%
\providecommand \@sanitize@url [0]{\catcode `\\12\catcode `\$12\catcode
  `\&12\catcode `\#12\catcode `\^12\catcode `\_12\catcode `\%12\relax}%
\providecommand \@@startlink[1]{}%
\providecommand \@@endlink[0]{}%
\providecommand \url  [0]{\begingroup\@sanitize@url \@url }%
\providecommand \@url [1]{\endgroup\@href {#1}{\urlprefix }}%
\providecommand \urlprefix  [0]{URL }%
\providecommand \Eprint [0]{\href }%
\providecommand \doibase [0]{http://dx.doi.org/}%
\providecommand \selectlanguage [0]{\@gobble}%
\providecommand \bibinfo  [0]{\@secondoftwo}%
\providecommand \bibfield  [0]{\@secondoftwo}%
\providecommand \translation [1]{[#1]}%
\providecommand \BibitemOpen [0]{}%
\providecommand \bibitemStop [0]{}%
\providecommand \bibitemNoStop [0]{.\EOS\space}%
\providecommand \EOS [0]{\spacefactor3000\relax}%
\providecommand \BibitemShut  [1]{\csname bibitem#1\endcsname}%
\let\auto@bib@innerbib\@empty
\bibitem [{\citenamefont {Kaplan}(2009)}]{Kaplan:2009yg}%
  \BibitemOpen
  \bibfield  {author} {\bibinfo {author} {\bibfnamefont {D.~B.}\ \bibnamefont
  {Kaplan}},\ }in\ \href@noop {} {\emph {\bibinfo {booktitle} {{Les Houches
  Summer School: Session 93: Modern perspectives in lattice QCD: Quantum field
  theory and high performance computing}}}}\ (\bibinfo {year} {2009})\ pp.\
  \bibinfo {pages} {223--272},\ \Eprint {http://arxiv.org/abs/0912.2560}
  {arXiv:0912.2560 [hep-lat]} \BibitemShut {NoStop}%
\bibitem [{\citenamefont {Schwinger}(1962)}]{PhysRev.128.2425}%
  \BibitemOpen
  \bibfield  {author} {\bibinfo {author} {\bibfnamefont {J.}~\bibnamefont
  {Schwinger}},\ }\href {\doibase 10.1103/PhysRev.128.2425} {\bibfield
  {journal} {\bibinfo  {journal} {Phys. Rev.}\ }\textbf {\bibinfo {volume}
  {128}},\ \bibinfo {pages} {2425} (\bibinfo {year} {1962})}\BibitemShut
  {NoStop}%
\bibitem [{\citenamefont {Adler}(1969)}]{Adler:1969gk}%
  \BibitemOpen
  \bibfield  {author} {\bibinfo {author} {\bibfnamefont {S.~L.}\ \bibnamefont
  {Adler}},\ }\href {\doibase 10.1103/PhysRev.177.2426} {\bibfield  {journal}
  {\bibinfo  {journal} {Phys. Rev.}\ }\textbf {\bibinfo {volume} {177}},\
  \bibinfo {pages} {2426} (\bibinfo {year} {1969})}\BibitemShut {NoStop}%
\bibitem [{\citenamefont {Bell}\ and\ \citenamefont
  {Jackiw}(1969)}]{Bell:1969ts}%
  \BibitemOpen
  \bibfield  {author} {\bibinfo {author} {\bibfnamefont {J.~S.}\ \bibnamefont
  {Bell}}\ and\ \bibinfo {author} {\bibfnamefont {R.}~\bibnamefont {Jackiw}},\
  }\href {\doibase 10.1007/BF02823296} {\bibfield  {journal} {\bibinfo
  {journal} {Nuovo Cim. A}\ }\textbf {\bibinfo {volume} {60}},\ \bibinfo
  {pages} {47} (\bibinfo {year} {1969})}\BibitemShut {NoStop}%
\bibitem [{\citenamefont {Kapustin}\ and\ \citenamefont
  {Thorngren}(2014)}]{Kapustin:2014lwa}%
  \BibitemOpen
  \bibfield  {author} {\bibinfo {author} {\bibfnamefont {A.}~\bibnamefont
  {Kapustin}}\ and\ \bibinfo {author} {\bibfnamefont {R.}~\bibnamefont
  {Thorngren}},\ }\href {\doibase 10.1103/PhysRevLett.112.231602} {\bibfield
  {journal} {\bibinfo  {journal} {Phys. Rev. Lett.}\ }\textbf {\bibinfo
  {volume} {112}},\ \bibinfo {pages} {231602} (\bibinfo {year} {2014})},\
  \Eprint {http://arxiv.org/abs/1403.0617} {arXiv:1403.0617 [hep-th]}
  \BibitemShut {NoStop}%
\bibitem [{\citenamefont {Seifnashri}(2024)}]{Seifnashri:2023dpa}%
  \BibitemOpen
  \bibfield  {author} {\bibinfo {author} {\bibfnamefont {S.}~\bibnamefont
  {Seifnashri}},\ }\href {\doibase 10.21468/SciPostPhys.16.4.098} {\bibfield
  {journal} {\bibinfo  {journal} {SciPost Phys.}\ }\textbf {\bibinfo {volume}
  {16}},\ \bibinfo {pages} {098} (\bibinfo {year} {2024})},\ \Eprint
  {http://arxiv.org/abs/2308.05151} {arXiv:2308.05151 [cond-mat.str-el]}
  \BibitemShut {NoStop}%
\bibitem [{\citenamefont {Sulejmanpasic}\ and\ \citenamefont
  {Gattringer}(2019)}]{Sulejmanpasic:2019ytl}%
  \BibitemOpen
  \bibfield  {author} {\bibinfo {author} {\bibfnamefont {T.}~\bibnamefont
  {Sulejmanpasic}}\ and\ \bibinfo {author} {\bibfnamefont {C.}~\bibnamefont
  {Gattringer}},\ }\href {\doibase 10.1016/j.nuclphysb.2019.114616} {\bibfield
  {journal} {\bibinfo  {journal} {Nucl. Phys. B}\ }\textbf {\bibinfo {volume}
  {943}},\ \bibinfo {pages} {114616} (\bibinfo {year} {2019})},\ \Eprint
  {http://arxiv.org/abs/1901.02637} {arXiv:1901.02637 [hep-lat]} \BibitemShut
  {NoStop}%
\bibitem [{\citenamefont {Gorantla}\ \emph {et~al.}(2021)\citenamefont
  {Gorantla}, \citenamefont {Lam}, \citenamefont {Seiberg},\ and\ \citenamefont
  {Shao}}]{Gorantla:2021svj}%
  \BibitemOpen
  \bibfield  {author} {\bibinfo {author} {\bibfnamefont {P.}~\bibnamefont
  {Gorantla}}, \bibinfo {author} {\bibfnamefont {H.~T.}\ \bibnamefont {Lam}},
  \bibinfo {author} {\bibfnamefont {N.}~\bibnamefont {Seiberg}}, \ and\
  \bibinfo {author} {\bibfnamefont {S.-H.}\ \bibnamefont {Shao}},\ }\href
  {\doibase 10.1063/5.0060808} {\bibfield  {journal} {\bibinfo  {journal} {J.
  Math. Phys.}\ }\textbf {\bibinfo {volume} {62}},\ \bibinfo {pages} {102301}
  (\bibinfo {year} {2021})},\ \Eprint {http://arxiv.org/abs/2103.01257}
  {arXiv:2103.01257 [cond-mat.str-el]} \BibitemShut {NoStop}%
\bibitem [{\citenamefont {Cheng}\ and\ \citenamefont
  {Seiberg}(2023)}]{Cheng:2022sgb}%
  \BibitemOpen
  \bibfield  {author} {\bibinfo {author} {\bibfnamefont {M.}~\bibnamefont
  {Cheng}}\ and\ \bibinfo {author} {\bibfnamefont {N.}~\bibnamefont
  {Seiberg}},\ }\href {\doibase 10.21468/SciPostPhys.15.2.051} {\bibfield
  {journal} {\bibinfo  {journal} {SciPost Phys.}\ }\textbf {\bibinfo {volume}
  {15}},\ \bibinfo {pages} {051} (\bibinfo {year} {2023})},\ \Eprint
  {http://arxiv.org/abs/2211.12543} {arXiv:2211.12543 [cond-mat.str-el]}
  \BibitemShut {NoStop}%
\bibitem [{\citenamefont {Fazza}\ and\ \citenamefont
  {Sulejmanpasic}(2023)}]{Fazza:2022fss}%
  \BibitemOpen
  \bibfield  {author} {\bibinfo {author} {\bibfnamefont {L.}~\bibnamefont
  {Fazza}}\ and\ \bibinfo {author} {\bibfnamefont {T.}~\bibnamefont
  {Sulejmanpasic}},\ }\href {\doibase 10.1007/JHEP05(2023)017} {\bibfield
  {journal} {\bibinfo  {journal} {JHEP}\ }\textbf {\bibinfo {volume} {05}},\
  \bibinfo {pages} {017} (\bibinfo {year} {2023})},\ \Eprint
  {http://arxiv.org/abs/2211.13047} {arXiv:2211.13047 [hep-th]} \BibitemShut
  {NoStop}%
\bibitem [{\citenamefont {Berkowitz}\ \emph {et~al.}(2024)\citenamefont
  {Berkowitz}, \citenamefont {Cherman},\ and\ \citenamefont
  {Jacobson}}]{Berkowitz:2023pnz}%
  \BibitemOpen
  \bibfield  {author} {\bibinfo {author} {\bibfnamefont {E.}~\bibnamefont
  {Berkowitz}}, \bibinfo {author} {\bibfnamefont {A.}~\bibnamefont {Cherman}},
  \ and\ \bibinfo {author} {\bibfnamefont {T.}~\bibnamefont {Jacobson}},\
  }\href {\doibase 10.1103/PhysRevD.110.014510} {\bibfield  {journal} {\bibinfo
   {journal} {Phys. Rev. D}\ }\textbf {\bibinfo {volume} {110}},\ \bibinfo
  {pages} {014510} (\bibinfo {year} {2024})},\ \Eprint
  {http://arxiv.org/abs/2310.17539} {arXiv:2310.17539 [hep-lat]} \BibitemShut
  {NoStop}%
\bibitem [{\citenamefont {Catterall}\ \emph {et~al.}(2018)\citenamefont
  {Catterall}, \citenamefont {Laiho},\ and\ \citenamefont
  {Unmuth-Yockey}}]{Catterall:2018lkj}%
  \BibitemOpen
  \bibfield  {author} {\bibinfo {author} {\bibfnamefont {S.}~\bibnamefont
  {Catterall}}, \bibinfo {author} {\bibfnamefont {J.}~\bibnamefont {Laiho}}, \
  and\ \bibinfo {author} {\bibfnamefont {J.}~\bibnamefont {Unmuth-Yockey}},\
  }\href {\doibase 10.1007/JHEP10(2018)013} {\bibfield  {journal} {\bibinfo
  {journal} {JHEP}\ }\textbf {\bibinfo {volume} {10}},\ \bibinfo {pages} {013}
  (\bibinfo {year} {2018})},\ \Eprint {http://arxiv.org/abs/1806.07845}
  {arXiv:1806.07845 [hep-lat]} \BibitemShut {NoStop}%
\bibitem [{\citenamefont {Schwinger}(1959)}]{Schwinger:1959xd}%
  \BibitemOpen
  \bibfield  {author} {\bibinfo {author} {\bibfnamefont {J.~S.}\ \bibnamefont
  {Schwinger}},\ }\href {\doibase 10.1103/PhysRevLett.3.296} {\bibfield
  {journal} {\bibinfo  {journal} {Phys. Rev. Lett.}\ }\textbf {\bibinfo
  {volume} {3}},\ \bibinfo {pages} {296} (\bibinfo {year} {1959})}\BibitemShut
  {NoStop}%
\bibitem [{\citenamefont {Treiman}\ \emph {et~al.}(2014)\citenamefont
  {Treiman}, \citenamefont {Witten}, \citenamefont {Jackiw},\ and\
  \citenamefont {Zumino}}]{Treiman:1986ep}%
  \BibitemOpen
  \bibfield  {author} {\bibinfo {author} {\bibfnamefont {S.~B.}\ \bibnamefont
  {Treiman}}, \bibinfo {author} {\bibfnamefont {E.}~\bibnamefont {Witten}},
  \bibinfo {author} {\bibfnamefont {R.}~\bibnamefont {Jackiw}}, \ and\ \bibinfo
  {author} {\bibfnamefont {B.}~\bibnamefont {Zumino}},\ }\href@noop {} {\emph
  {\bibinfo {title} {{Current Algebra and Anomalies}}}}\ (\bibinfo  {publisher}
  {Princeton University Press},\ \bibinfo {year} {2014})\BibitemShut {NoStop}%
\bibitem [{\citenamefont {Kapustin}\ and\ \citenamefont
  {Sopenko}(2024)}]{Kapustin:2024rrm}%
  \BibitemOpen
  \bibfield  {author} {\bibinfo {author} {\bibfnamefont {A.}~\bibnamefont
  {Kapustin}}\ and\ \bibinfo {author} {\bibfnamefont {N.}~\bibnamefont
  {Sopenko}},\ }\href@noop {} {\  (\bibinfo {year} {2024})},\ \Eprint
  {http://arxiv.org/abs/2401.02533} {arXiv:2401.02533 [math-ph]} \BibitemShut
  {NoStop}%
\bibitem [{\citenamefont {Nielsen}\ and\ \citenamefont
  {Ninomiya}(1981{\natexlab{a}})}]{Nielsen:1980rz}%
  \BibitemOpen
  \bibfield  {author} {\bibinfo {author} {\bibfnamefont {H.~B.}\ \bibnamefont
  {Nielsen}}\ and\ \bibinfo {author} {\bibfnamefont {M.}~\bibnamefont
  {Ninomiya}},\ }\href {\doibase 10.1016/0550-3213(82)90011-6} {\bibfield
  {journal} {\bibinfo  {journal} {Nucl. Phys. B}\ }\textbf {\bibinfo {volume}
  {185}},\ \bibinfo {pages} {20} (\bibinfo {year} {1981}{\natexlab{a}})},\
  \bibinfo {note} {[Erratum: Nucl.Phys.B 195, 541 (1982)]}\BibitemShut
  {NoStop}%
\bibitem [{\citenamefont {Nielsen}\ and\ \citenamefont
  {Ninomiya}(1981{\natexlab{b}})}]{Nielsen:1981hk}%
  \BibitemOpen
  \bibfield  {author} {\bibinfo {author} {\bibfnamefont {H.~B.}\ \bibnamefont
  {Nielsen}}\ and\ \bibinfo {author} {\bibfnamefont {M.}~\bibnamefont
  {Ninomiya}},\ }\href {\doibase 10.1016/0370-2693(81)91026-1} {\bibfield
  {journal} {\bibinfo  {journal} {Phys. Lett. B}\ }\textbf {\bibinfo {volume}
  {105}},\ \bibinfo {pages} {219} (\bibinfo {year}
  {1981}{\natexlab{b}})}\BibitemShut {NoStop}%
\bibitem [{\citenamefont {Nielsen}\ and\ \citenamefont
  {Ninomiya}(1981{\natexlab{c}})}]{Nielsen:1981xu}%
  \BibitemOpen
  \bibfield  {author} {\bibinfo {author} {\bibfnamefont {H.~B.}\ \bibnamefont
  {Nielsen}}\ and\ \bibinfo {author} {\bibfnamefont {M.}~\bibnamefont
  {Ninomiya}},\ }\href {\doibase 10.1016/0550-3213(81)90524-1} {\bibfield
  {journal} {\bibinfo  {journal} {Nucl. Phys. B}\ }\textbf {\bibinfo {volume}
  {193}},\ \bibinfo {pages} {173} (\bibinfo {year}
  {1981}{\natexlab{c}})}\BibitemShut {NoStop}%
\bibitem [{\citenamefont {Friedan}(1982)}]{Friedan:1982nk}%
  \BibitemOpen
  \bibfield  {author} {\bibinfo {author} {\bibfnamefont {D.}~\bibnamefont
  {Friedan}},\ }\href {\doibase 10.1007/BF01403500} {\bibfield  {journal}
  {\bibinfo  {journal} {Commun. Math. Phys.}\ }\textbf {\bibinfo {volume}
  {85}},\ \bibinfo {pages} {481} (\bibinfo {year} {1982})}\BibitemShut
  {NoStop}%
\bibitem [{\citenamefont {Onsager}(1944)}]{PhysRev.65.117}%
  \BibitemOpen
  \bibfield  {author} {\bibinfo {author} {\bibfnamefont {L.}~\bibnamefont
  {Onsager}},\ }\href {\doibase 10.1103/PhysRev.65.117} {\bibfield  {journal}
  {\bibinfo  {journal} {Phys. Rev.}\ }\textbf {\bibinfo {volume} {65}},\
  \bibinfo {pages} {117} (\bibinfo {year} {1944})}\BibitemShut {NoStop}%
\bibitem [{\citenamefont {Johnson}(1963)}]{JOHNSON1963253}%
  \BibitemOpen
  \bibfield  {author} {\bibinfo {author} {\bibfnamefont {K.}~\bibnamefont
  {Johnson}},\ }\href {\doibase https://doi.org/10.1016/S0375-9601(63)95573-7}
  {\bibfield  {journal} {\bibinfo  {journal} {Physics Letters}\ }\textbf
  {\bibinfo {volume} {5}},\ \bibinfo {pages} {253} (\bibinfo {year}
  {1963})}\BibitemShut {NoStop}%
\bibitem [{\citenamefont {Kogut}\ and\ \citenamefont
  {Susskind}(1975)}]{Kogut:1974ag}%
  \BibitemOpen
  \bibfield  {author} {\bibinfo {author} {\bibfnamefont {J.~B.}\ \bibnamefont
  {Kogut}}\ and\ \bibinfo {author} {\bibfnamefont {L.}~\bibnamefont
  {Susskind}},\ }\href {\doibase 10.1103/PhysRevD.11.395} {\bibfield  {journal}
  {\bibinfo  {journal} {Phys. Rev. D}\ }\textbf {\bibinfo {volume} {11}},\
  \bibinfo {pages} {395} (\bibinfo {year} {1975})}\BibitemShut {NoStop}%
\bibitem [{\citenamefont {Banks}\ \emph {et~al.}(1976)\citenamefont {Banks},
  \citenamefont {Susskind},\ and\ \citenamefont {Kogut}}]{Banks:1975gq}%
  \BibitemOpen
  \bibfield  {author} {\bibinfo {author} {\bibfnamefont {T.}~\bibnamefont
  {Banks}}, \bibinfo {author} {\bibfnamefont {L.}~\bibnamefont {Susskind}}, \
  and\ \bibinfo {author} {\bibfnamefont {J.~B.}\ \bibnamefont {Kogut}},\ }\href
  {\doibase 10.1103/PhysRevD.13.1043} {\bibfield  {journal} {\bibinfo
  {journal} {Phys. Rev. D}\ }\textbf {\bibinfo {volume} {13}},\ \bibinfo
  {pages} {1043} (\bibinfo {year} {1976})}\BibitemShut {NoStop}%
\bibitem [{\citenamefont {Susskind}(1977)}]{Susskind:1976jm}%
  \BibitemOpen
  \bibfield  {author} {\bibinfo {author} {\bibfnamefont {L.}~\bibnamefont
  {Susskind}},\ }\href {\doibase 10.1103/PhysRevD.16.3031} {\bibfield
  {journal} {\bibinfo  {journal} {Phys. Rev. D}\ }\textbf {\bibinfo {volume}
  {16}},\ \bibinfo {pages} {3031} (\bibinfo {year} {1977})}\BibitemShut
  {NoStop}%
\bibitem [{\citenamefont {Seiberg}\ and\ \citenamefont
  {Shao}(2024)}]{Seiberg:2023cdc}%
  \BibitemOpen
  \bibfield  {author} {\bibinfo {author} {\bibfnamefont {N.}~\bibnamefont
  {Seiberg}}\ and\ \bibinfo {author} {\bibfnamefont {S.-H.}\ \bibnamefont
  {Shao}},\ }\href {\doibase 10.21468/SciPostPhys.16.3.064} {\bibfield
  {journal} {\bibinfo  {journal} {SciPost Phys.}\ }\textbf {\bibinfo {volume}
  {16}},\ \bibinfo {pages} {064} (\bibinfo {year} {2024})},\ \Eprint
  {http://arxiv.org/abs/2307.02534} {arXiv:2307.02534 [cond-mat.str-el]}
  \BibitemShut {NoStop}%
\bibitem [{\citenamefont {Clancy}\ \emph {et~al.}(2024)\citenamefont {Clancy},
  \citenamefont {Kaplan},\ and\ \citenamefont {Singh}}]{Clancy:2023ino}%
  \BibitemOpen
  \bibfield  {author} {\bibinfo {author} {\bibfnamefont {M.}~\bibnamefont
  {Clancy}}, \bibinfo {author} {\bibfnamefont {D.~B.}\ \bibnamefont {Kaplan}},
  \ and\ \bibinfo {author} {\bibfnamefont {H.}~\bibnamefont {Singh}},\ }\href
  {\doibase 10.1103/PhysRevD.109.014502} {\bibfield  {journal} {\bibinfo
  {journal} {Phys. Rev. D}\ }\textbf {\bibinfo {volume} {109}},\ \bibinfo
  {pages} {014502} (\bibinfo {year} {2024})},\ \Eprint
  {http://arxiv.org/abs/2309.08542} {arXiv:2309.08542 [hep-lat]} \BibitemShut
  {NoStop}%
\bibitem [{\citenamefont {Thacker}(1995)}]{Thacker:1994ns}%
  \BibitemOpen
  \bibfield  {author} {\bibinfo {author} {\bibfnamefont {H.~B.}\ \bibnamefont
  {Thacker}},\ }\href {\doibase 10.1016/0920-5632(95)00337-9} {\bibfield
  {journal} {\bibinfo  {journal} {Nucl. Phys. B Proc. Suppl.}\ }\textbf
  {\bibinfo {volume} {42}},\ \bibinfo {pages} {642} (\bibinfo {year} {1995})},\
  \Eprint {http://arxiv.org/abs/hep-lat/9412019} {arXiv:hep-lat/9412019}
  \BibitemShut {NoStop}%
\bibitem [{\citenamefont {Horvath}\ and\ \citenamefont
  {Thacker}(1999)}]{Horvath:1998gq}%
  \BibitemOpen
  \bibfield  {author} {\bibinfo {author} {\bibfnamefont {I.}~\bibnamefont
  {Horvath}}\ and\ \bibinfo {author} {\bibfnamefont {H.~B.}\ \bibnamefont
  {Thacker}},\ }\href {\doibase 10.1016/S0920-5632(99)85172-X} {\bibfield
  {journal} {\bibinfo  {journal} {Nucl. Phys. B Proc. Suppl.}\ }\textbf
  {\bibinfo {volume} {73}},\ \bibinfo {pages} {682} (\bibinfo {year} {1999})},\
  \Eprint {http://arxiv.org/abs/hep-lat/9809108} {arXiv:hep-lat/9809108}
  \BibitemShut {NoStop}%
\bibitem [{\citenamefont {Banks}\ \emph {et~al.}()\citenamefont {Banks},
  \citenamefont {Seiberg},\ and\ \citenamefont {Shao}}]{SS}%
  \BibitemOpen
  \bibfield  {author} {\bibinfo {author} {\bibfnamefont {T.}~\bibnamefont
  {Banks}}, \bibinfo {author} {\bibfnamefont {N.}~\bibnamefont {Seiberg}}, \
  and\ \bibinfo {author} {\bibfnamefont {S.-H.}\ \bibnamefont {Shao}},\
  }\href@noop {} {\bibinfo  {journal} {unpublished}\ }\BibitemShut {NoStop}%
\bibitem [{\citenamefont {Golterman}(2024)}]{Golterman:2024xos}%
  \BibitemOpen
\bibfield  {journal} {  }\bibfield  {author} {\bibinfo {author} {\bibfnamefont
  {M.}~\bibnamefont {Golterman}},\ }\href@noop {} {\enquote {\bibinfo {title}
  {{Staggered fermions}},}\ } (\bibinfo {year} {2024}),\ \Eprint
  {http://arxiv.org/abs/2406.02906} {arXiv:2406.02906 [hep-lat]} \BibitemShut
  {NoStop}%
\bibitem [{\citenamefont {Karsten}\ and\ \citenamefont
  {Smit}(1981)}]{Karsten:1980wd}%
  \BibitemOpen
  \bibfield  {author} {\bibinfo {author} {\bibfnamefont {L.~H.}\ \bibnamefont
  {Karsten}}\ and\ \bibinfo {author} {\bibfnamefont {J.}~\bibnamefont {Smit}},\
  }\href {\doibase 10.1016/0550-3213(81)90549-6} {\bibfield  {journal}
  {\bibinfo  {journal} {Nucl. Phys. B}\ }\textbf {\bibinfo {volume} {183}},\
  \bibinfo {pages} {103} (\bibinfo {year} {1981})}\BibitemShut {NoStop}%
\bibitem [{\citenamefont {Golterman}\ and\ \citenamefont
  {Smit}(1984)}]{Golterman:1984cy}%
  \BibitemOpen
  \bibfield  {author} {\bibinfo {author} {\bibfnamefont {M.~F.~L.}\
  \bibnamefont {Golterman}}\ and\ \bibinfo {author} {\bibfnamefont
  {J.}~\bibnamefont {Smit}},\ }\href {\doibase 10.1016/0550-3213(84)90424-3}
  {\bibfield  {journal} {\bibinfo  {journal} {Nucl. Phys. B}\ }\textbf
  {\bibinfo {volume} {245}},\ \bibinfo {pages} {61} (\bibinfo {year}
  {1984})}\BibitemShut {NoStop}%
\bibitem [{\citenamefont {Catterall}(2023)}]{Catterall:2022jky}%
  \BibitemOpen
  \bibfield  {author} {\bibinfo {author} {\bibfnamefont {S.}~\bibnamefont
  {Catterall}},\ }\href {\doibase 10.1103/PhysRevD.107.014501} {\bibfield
  {journal} {\bibinfo  {journal} {Phys. Rev. D}\ }\textbf {\bibinfo {volume}
  {107}},\ \bibinfo {pages} {014501} (\bibinfo {year} {2023})},\ \Eprint
  {http://arxiv.org/abs/2209.03828} {arXiv:2209.03828 [hep-lat]} \BibitemShut
  {NoStop}%
\bibitem [{\citenamefont {Drell}\ \emph {et~al.}(1976)\citenamefont {Drell},
  \citenamefont {Weinstein},\ and\ \citenamefont
  {Yankielowicz}}]{Drell:1976mj}%
  \BibitemOpen
  \bibfield  {author} {\bibinfo {author} {\bibfnamefont {S.~D.}\ \bibnamefont
  {Drell}}, \bibinfo {author} {\bibfnamefont {M.}~\bibnamefont {Weinstein}}, \
  and\ \bibinfo {author} {\bibfnamefont {S.}~\bibnamefont {Yankielowicz}},\
  }\href {\doibase 10.1103/PhysRevD.14.1627} {\bibfield  {journal} {\bibinfo
  {journal} {Phys. Rev. D}\ }\textbf {\bibinfo {volume} {14}},\ \bibinfo
  {pages} {1627} (\bibinfo {year} {1976})}\BibitemShut {NoStop}%
\bibitem [{\citenamefont {Karsten}\ and\ \citenamefont
  {Smit}(1978)}]{Karsten:1978nb}%
  \BibitemOpen
  \bibfield  {author} {\bibinfo {author} {\bibfnamefont {L.~H.}\ \bibnamefont
  {Karsten}}\ and\ \bibinfo {author} {\bibfnamefont {J.}~\bibnamefont {Smit}},\
  }\href {\doibase 10.1016/0550-3213(78)90385-1} {\bibfield  {journal}
  {\bibinfo  {journal} {Nucl. Phys. B}\ }\textbf {\bibinfo {volume} {144}},\
  \bibinfo {pages} {536} (\bibinfo {year} {1978})}\BibitemShut {NoStop}%
\bibitem [{\citenamefont {Weinstein}(1982)}]{Weinstein:1982ht}%
  \BibitemOpen
  \bibfield  {author} {\bibinfo {author} {\bibfnamefont {M.}~\bibnamefont
  {Weinstein}},\ }\href {\doibase 10.1103/PhysRevD.26.839} {\bibfield
  {journal} {\bibinfo  {journal} {Phys. Rev. D}\ }\textbf {\bibinfo {volume}
  {26}},\ \bibinfo {pages} {839} (\bibinfo {year} {1982})}\BibitemShut
  {NoStop}%
\bibitem [{\citenamefont {Creutz}\ \emph {et~al.}(2002)\citenamefont {Creutz},
  \citenamefont {Horvath},\ and\ \citenamefont {Neuberger}}]{Creutz:2001wp}%
  \BibitemOpen
  \bibfield  {author} {\bibinfo {author} {\bibfnamefont {M.}~\bibnamefont
  {Creutz}}, \bibinfo {author} {\bibfnamefont {I.}~\bibnamefont {Horvath}}, \
  and\ \bibinfo {author} {\bibfnamefont {H.}~\bibnamefont {Neuberger}},\ }\href
  {\doibase 10.1016/S0920-5632(01)01836-9} {\bibfield  {journal} {\bibinfo
  {journal} {Nucl. Phys. B Proc. Suppl.}\ }\textbf {\bibinfo {volume} {106}},\
  \bibinfo {pages} {760} (\bibinfo {year} {2002})},\ \Eprint
  {http://arxiv.org/abs/hep-lat/0110009} {arXiv:hep-lat/0110009} \BibitemShut
  {NoStop}%
\bibitem [{\citenamefont {Narayanan}\ and\ \citenamefont
  {Neuberger}(1995)}]{Narayanan:1994gw}%
  \BibitemOpen
  \bibfield  {author} {\bibinfo {author} {\bibfnamefont {R.}~\bibnamefont
  {Narayanan}}\ and\ \bibinfo {author} {\bibfnamefont {H.}~\bibnamefont
  {Neuberger}},\ }\href {\doibase 10.1016/0550-3213(95)00111-5} {\bibfield
  {journal} {\bibinfo  {journal} {Nucl. Phys. B}\ }\textbf {\bibinfo {volume}
  {443}},\ \bibinfo {pages} {305} (\bibinfo {year} {1995})},\ \Eprint
  {http://arxiv.org/abs/hep-th/9411108} {arXiv:hep-th/9411108} \BibitemShut
  {NoStop}%
\bibitem [{\citenamefont {Neuberger}(1998{\natexlab{a}})}]{Neuberger:1997fp}%
  \BibitemOpen
  \bibfield  {author} {\bibinfo {author} {\bibfnamefont {H.}~\bibnamefont
  {Neuberger}},\ }\href {\doibase 10.1016/S0370-2693(97)01368-3} {\bibfield
  {journal} {\bibinfo  {journal} {Phys. Lett. B}\ }\textbf {\bibinfo {volume}
  {417}},\ \bibinfo {pages} {141} (\bibinfo {year} {1998}{\natexlab{a}})},\
  \Eprint {http://arxiv.org/abs/hep-lat/9707022} {arXiv:hep-lat/9707022}
  \BibitemShut {NoStop}%
\bibitem [{\citenamefont {Neuberger}(1998{\natexlab{b}})}]{Neuberger:1998wv}%
  \BibitemOpen
  \bibfield  {author} {\bibinfo {author} {\bibfnamefont {H.}~\bibnamefont
  {Neuberger}},\ }\href {\doibase 10.1016/S0370-2693(98)00355-4} {\bibfield
  {journal} {\bibinfo  {journal} {Phys. Lett. B}\ }\textbf {\bibinfo {volume}
  {427}},\ \bibinfo {pages} {353} (\bibinfo {year} {1998}{\natexlab{b}})},\
  \Eprint {http://arxiv.org/abs/hep-lat/9801031} {arXiv:hep-lat/9801031}
  \BibitemShut {NoStop}%
\bibitem [{\citenamefont {Ginsparg}\ and\ \citenamefont
  {Wilson}(1982)}]{PhysRevD.25.2649}%
  \BibitemOpen
  \bibfield  {author} {\bibinfo {author} {\bibfnamefont {P.~H.}\ \bibnamefont
  {Ginsparg}}\ and\ \bibinfo {author} {\bibfnamefont {K.~G.}\ \bibnamefont
  {Wilson}},\ }\href {\doibase 10.1103/PhysRevD.25.2649} {\bibfield  {journal}
  {\bibinfo  {journal} {Phys. Rev. D}\ }\textbf {\bibinfo {volume} {25}},\
  \bibinfo {pages} {2649} (\bibinfo {year} {1982})}\BibitemShut {NoStop}%
\bibitem [{\citenamefont {Kaplan}(1992)}]{Kaplan:1992bt}%
  \BibitemOpen
  \bibfield  {author} {\bibinfo {author} {\bibfnamefont {D.~B.}\ \bibnamefont
  {Kaplan}},\ }\href {\doibase 10.1016/0370-2693(92)91112-M} {\bibfield
  {journal} {\bibinfo  {journal} {Phys. Lett. B}\ }\textbf {\bibinfo {volume}
  {288}},\ \bibinfo {pages} {342} (\bibinfo {year} {1992})},\ \Eprint
  {http://arxiv.org/abs/hep-lat/9206013} {arXiv:hep-lat/9206013} \BibitemShut
  {NoStop}%
\bibitem [{\citenamefont {Kaplan}\ and\ \citenamefont
  {Sen}(2024)}]{Kaplan:2023pvd}%
  \BibitemOpen
  \bibfield  {author} {\bibinfo {author} {\bibfnamefont {D.~B.}\ \bibnamefont
  {Kaplan}}\ and\ \bibinfo {author} {\bibfnamefont {S.}~\bibnamefont {Sen}},\
  }\href {\doibase 10.1103/PhysRevLett.132.141604} {\bibfield  {journal}
  {\bibinfo  {journal} {Phys. Rev. Lett.}\ }\textbf {\bibinfo {volume} {132}},\
  \bibinfo {pages} {141604} (\bibinfo {year} {2024})},\ \Eprint
  {http://arxiv.org/abs/2312.04012} {arXiv:2312.04012 [hep-lat]} \BibitemShut
  {NoStop}%
\bibitem [{\citenamefont {Kaplan}(2024)}]{Kaplan:2023pxd}%
  \BibitemOpen
  \bibfield  {author} {\bibinfo {author} {\bibfnamefont {D.~B.}\ \bibnamefont
  {Kaplan}},\ }\href {\doibase 10.1103/PhysRevLett.132.141603} {\bibfield
  {journal} {\bibinfo  {journal} {Phys. Rev. Lett.}\ }\textbf {\bibinfo
  {volume} {132}},\ \bibinfo {pages} {141603} (\bibinfo {year} {2024})},\
  \Eprint {http://arxiv.org/abs/2312.01494} {arXiv:2312.01494 [hep-lat]}
  \BibitemShut {NoStop}%
\bibitem [{\citenamefont {Carroll}\ \emph {et~al.}(1976)\citenamefont
  {Carroll}, \citenamefont {Kogut}, \citenamefont {Sinclair},\ and\
  \citenamefont {Susskind}}]{Carroll:1975gb}%
  \BibitemOpen
  \bibfield  {author} {\bibinfo {author} {\bibfnamefont {A.}~\bibnamefont
  {Carroll}}, \bibinfo {author} {\bibfnamefont {J.~B.}\ \bibnamefont {Kogut}},
  \bibinfo {author} {\bibfnamefont {D.~K.}\ \bibnamefont {Sinclair}}, \ and\
  \bibinfo {author} {\bibfnamefont {L.}~\bibnamefont {Susskind}},\ }\href
  {\doibase 10.1103/PhysRevD.13.2270} {\bibfield  {journal} {\bibinfo
  {journal} {Phys. Rev. D}\ }\textbf {\bibinfo {volume} {13}},\ \bibinfo
  {pages} {2270} (\bibinfo {year} {1976})},\ \bibinfo {note} {[Erratum:
  Phys.Rev.D 14, 1729 (1976)]}\BibitemShut {NoStop}%
\bibitem [{\citenamefont {Hamer}\ \emph {et~al.}(1997)\citenamefont {Hamer},
  \citenamefont {Zheng},\ and\ \citenamefont {Oitmaa}}]{Hamer:1997dx}%
  \BibitemOpen
  \bibfield  {author} {\bibinfo {author} {\bibfnamefont {C.~J.}\ \bibnamefont
  {Hamer}}, \bibinfo {author} {\bibfnamefont {W.-h.}\ \bibnamefont {Zheng}}, \
  and\ \bibinfo {author} {\bibfnamefont {J.}~\bibnamefont {Oitmaa}},\ }\href
  {\doibase 10.1103/PhysRevD.56.55} {\bibfield  {journal} {\bibinfo  {journal}
  {Phys. Rev. D}\ }\textbf {\bibinfo {volume} {56}},\ \bibinfo {pages} {55}
  (\bibinfo {year} {1997})},\ \Eprint {http://arxiv.org/abs/hep-lat/9701015}
  {arXiv:hep-lat/9701015} \BibitemShut {NoStop}%
\bibitem [{\citenamefont {Dempsey}\ \emph {et~al.}(2022)\citenamefont
  {Dempsey}, \citenamefont {Klebanov}, \citenamefont {Pufu},\ and\
  \citenamefont {Zan}}]{Dempsey:2022nys}%
  \BibitemOpen
  \bibfield  {author} {\bibinfo {author} {\bibfnamefont {R.}~\bibnamefont
  {Dempsey}}, \bibinfo {author} {\bibfnamefont {I.~R.}\ \bibnamefont
  {Klebanov}}, \bibinfo {author} {\bibfnamefont {S.~S.}\ \bibnamefont {Pufu}},
  \ and\ \bibinfo {author} {\bibfnamefont {B.}~\bibnamefont {Zan}},\ }\href
  {\doibase 10.1103/PhysRevResearch.4.043133} {\bibfield  {journal} {\bibinfo
  {journal} {Phys. Rev. Res.}\ }\textbf {\bibinfo {volume} {4}},\ \bibinfo
  {pages} {043133} (\bibinfo {year} {2022})},\ \Eprint
  {http://arxiv.org/abs/2206.05308} {arXiv:2206.05308 [hep-th]} \BibitemShut
  {NoStop}%
\bibitem [{\citenamefont {Shifman}(1991)}]{SHIFMAN1991341}%
  \BibitemOpen
  \bibfield  {author} {\bibinfo {author} {\bibfnamefont {M.}~\bibnamefont
  {Shifman}},\ }\href {\doibase https://doi.org/10.1016/0370-1573(91)90020-M}
  {\bibfield  {journal} {\bibinfo  {journal} {Physics Reports}\ }\textbf
  {\bibinfo {volume} {209}},\ \bibinfo {pages} {341} (\bibinfo {year}
  {1991})}\BibitemShut {NoStop}%
\bibitem [{\citenamefont {Manton}(1985)}]{Manton:1985jm}%
  \BibitemOpen
  \bibfield  {author} {\bibinfo {author} {\bibfnamefont {N.~S.}\ \bibnamefont
  {Manton}},\ }\href {\doibase 10.1016/0003-4916(85)90199-X} {\bibfield
  {journal} {\bibinfo  {journal} {Annals Phys.}\ }\textbf {\bibinfo {volume}
  {159}},\ \bibinfo {pages} {220} (\bibinfo {year} {1985})}\BibitemShut
  {NoStop}%
\bibitem [{\citenamefont {Chang}\ \emph {et~al.}(2019)\citenamefont {Chang},
  \citenamefont {Lin}, \citenamefont {Shao}, \citenamefont {Wang},\ and\
  \citenamefont {Yin}}]{Chang:2018iay}%
  \BibitemOpen
  \bibfield  {author} {\bibinfo {author} {\bibfnamefont {C.-M.}\ \bibnamefont
  {Chang}}, \bibinfo {author} {\bibfnamefont {Y.-H.}\ \bibnamefont {Lin}},
  \bibinfo {author} {\bibfnamefont {S.-H.}\ \bibnamefont {Shao}}, \bibinfo
  {author} {\bibfnamefont {Y.}~\bibnamefont {Wang}}, \ and\ \bibinfo {author}
  {\bibfnamefont {X.}~\bibnamefont {Yin}},\ }\href {\doibase
  10.1007/JHEP01(2019)026} {\bibfield  {journal} {\bibinfo  {journal} {JHEP}\
  }\textbf {\bibinfo {volume} {01}},\ \bibinfo {pages} {026} (\bibinfo {year}
  {2019})},\ \Eprint {http://arxiv.org/abs/1802.04445} {arXiv:1802.04445
  [hep-th]} \BibitemShut {NoStop}%
\bibitem [{\citenamefont {Wen}(2019)}]{Wen:2018zux}%
  \BibitemOpen
  \bibfield  {author} {\bibinfo {author} {\bibfnamefont {X.-G.}\ \bibnamefont
  {Wen}},\ }\href {\doibase 10.1103/PhysRevB.99.205139} {\bibfield  {journal}
  {\bibinfo  {journal} {Phys. Rev. B}\ }\textbf {\bibinfo {volume} {99}},\
  \bibinfo {pages} {205139} (\bibinfo {year} {2019})},\ \Eprint
  {http://arxiv.org/abs/1812.02517} {arXiv:1812.02517 [cond-mat.str-el]}
  \BibitemShut {NoStop}%
\bibitem [{\citenamefont {Thorngren}\ and\ \citenamefont
  {Wang}(2024{\natexlab{a}})}]{Thorngren:2019iar}%
  \BibitemOpen
  \bibfield  {author} {\bibinfo {author} {\bibfnamefont {R.}~\bibnamefont
  {Thorngren}}\ and\ \bibinfo {author} {\bibfnamefont {Y.}~\bibnamefont
  {Wang}},\ }\href {\doibase 10.1007/JHEP04(2024)132} {\bibfield  {journal}
  {\bibinfo  {journal} {JHEP}\ }\textbf {\bibinfo {volume} {04}},\ \bibinfo
  {pages} {132} (\bibinfo {year} {2024}{\natexlab{a}})},\ \Eprint
  {http://arxiv.org/abs/1912.02817} {arXiv:1912.02817 [hep-th]} \BibitemShut
  {NoStop}%
\bibitem [{\citenamefont {Choi}\ \emph {et~al.}(2022)\citenamefont {Choi},
  \citenamefont {Cordova}, \citenamefont {Hsin}, \citenamefont {Lam},\ and\
  \citenamefont {Shao}}]{Choi:2021kmx}%
  \BibitemOpen
  \bibfield  {author} {\bibinfo {author} {\bibfnamefont {Y.}~\bibnamefont
  {Choi}}, \bibinfo {author} {\bibfnamefont {C.}~\bibnamefont {Cordova}},
  \bibinfo {author} {\bibfnamefont {P.-S.}\ \bibnamefont {Hsin}}, \bibinfo
  {author} {\bibfnamefont {H.~T.}\ \bibnamefont {Lam}}, \ and\ \bibinfo
  {author} {\bibfnamefont {S.-H.}\ \bibnamefont {Shao}},\ }\href {\doibase
  10.1103/PhysRevD.105.125016} {\bibfield  {journal} {\bibinfo  {journal}
  {Phys. Rev. D}\ }\textbf {\bibinfo {volume} {105}},\ \bibinfo {pages}
  {125016} (\bibinfo {year} {2022})},\ \Eprint
  {http://arxiv.org/abs/2111.01139} {arXiv:2111.01139 [hep-th]} \BibitemShut
  {NoStop}%
\bibitem [{\citenamefont {Komargodski}\ \emph {et~al.}(2021)\citenamefont
  {Komargodski}, \citenamefont {Ohmori}, \citenamefont {Roumpedakis},\ and\
  \citenamefont {Seifnashri}}]{Komargodski:2020mxz}%
  \BibitemOpen
  \bibfield  {author} {\bibinfo {author} {\bibfnamefont {Z.}~\bibnamefont
  {Komargodski}}, \bibinfo {author} {\bibfnamefont {K.}~\bibnamefont {Ohmori}},
  \bibinfo {author} {\bibfnamefont {K.}~\bibnamefont {Roumpedakis}}, \ and\
  \bibinfo {author} {\bibfnamefont {S.}~\bibnamefont {Seifnashri}},\ }\href
  {\doibase 10.1007/JHEP03(2021)103} {\bibfield  {journal} {\bibinfo  {journal}
  {JHEP}\ }\textbf {\bibinfo {volume} {03}},\ \bibinfo {pages} {103} (\bibinfo
  {year} {2021})},\ \Eprint {http://arxiv.org/abs/2008.07567} {arXiv:2008.07567
  [hep-th]} \BibitemShut {NoStop}%
\bibitem [{\citenamefont {Choi}\ \emph {et~al.}(2023)\citenamefont {Choi},
  \citenamefont {Rayhaun}, \citenamefont {Sanghavi},\ and\ \citenamefont
  {Shao}}]{Choi:2023xjw}%
  \BibitemOpen
  \bibfield  {author} {\bibinfo {author} {\bibfnamefont {Y.}~\bibnamefont
  {Choi}}, \bibinfo {author} {\bibfnamefont {B.~C.}\ \bibnamefont {Rayhaun}},
  \bibinfo {author} {\bibfnamefont {Y.}~\bibnamefont {Sanghavi}}, \ and\
  \bibinfo {author} {\bibfnamefont {S.-H.}\ \bibnamefont {Shao}},\ }\href
  {\doibase 10.1103/PhysRevD.108.125005} {\bibfield  {journal} {\bibinfo
  {journal} {Phys. Rev. D}\ }\textbf {\bibinfo {volume} {108}},\ \bibinfo
  {pages} {125005} (\bibinfo {year} {2023})},\ \Eprint
  {http://arxiv.org/abs/2305.09713} {arXiv:2305.09713 [hep-th]} \BibitemShut
  {NoStop}%
\bibitem [{\citenamefont {Choi}\ \emph {et~al.}(2024)\citenamefont {Choi},
  \citenamefont {Lu},\ and\ \citenamefont {Sun}}]{Choi:2023vgk}%
  \BibitemOpen
  \bibfield  {author} {\bibinfo {author} {\bibfnamefont {Y.}~\bibnamefont
  {Choi}}, \bibinfo {author} {\bibfnamefont {D.-C.}\ \bibnamefont {Lu}}, \ and\
  \bibinfo {author} {\bibfnamefont {Z.}~\bibnamefont {Sun}},\ }\href {\doibase
  10.1007/JHEP01(2024)142} {\bibfield  {journal} {\bibinfo  {journal} {JHEP}\
  }\textbf {\bibinfo {volume} {01}},\ \bibinfo {pages} {142} (\bibinfo {year}
  {2024})},\ \Eprint {http://arxiv.org/abs/2310.19867} {arXiv:2310.19867
  [hep-th]} \BibitemShut {NoStop}%
\bibitem [{\citenamefont {Lieb}\ \emph {et~al.}(1961)\citenamefont {Lieb},
  \citenamefont {Schultz},\ and\ \citenamefont {Mattis}}]{LIEB1961407}%
  \BibitemOpen
  \bibfield  {author} {\bibinfo {author} {\bibfnamefont {E.}~\bibnamefont
  {Lieb}}, \bibinfo {author} {\bibfnamefont {T.}~\bibnamefont {Schultz}}, \
  and\ \bibinfo {author} {\bibfnamefont {D.}~\bibnamefont {Mattis}},\ }\href
  {\doibase https://doi.org/10.1016/0003-4916(61)90115-4} {\bibfield  {journal}
  {\bibinfo  {journal} {Annals of Physics}\ }\textbf {\bibinfo {volume} {16}},\
  \bibinfo {pages} {407} (\bibinfo {year} {1961})}\BibitemShut {NoStop}%
\bibitem [{\citenamefont {{Oshikawa}}(2000)}]{2000PhRvL..84.3370O}%
  \BibitemOpen
  \bibfield  {author} {\bibinfo {author} {\bibfnamefont {M.}~\bibnamefont
  {{Oshikawa}}},\ }\href {\doibase 10.1103/PhysRevLett.84.3370} {\bibfield
  {journal} {\bibinfo  {journal} {\prl}\ }\textbf {\bibinfo {volume} {84}},\
  \bibinfo {pages} {3370} (\bibinfo {year} {2000})},\ \Eprint
  {http://arxiv.org/abs/cond-mat/0002392} {arXiv:cond-mat/0002392
  [cond-mat.str-el]} \BibitemShut {NoStop}%
\bibitem [{\citenamefont {Hastings}(2004)}]{Hastings:2003zx}%
  \BibitemOpen
  \bibfield  {author} {\bibinfo {author} {\bibfnamefont {M.~B.}\ \bibnamefont
  {Hastings}},\ }\href {\doibase 10.1103/PhysRevB.69.104431} {\bibfield
  {journal} {\bibinfo  {journal} {Phys. Rev. B}\ }\textbf {\bibinfo {volume}
  {69}},\ \bibinfo {pages} {104431} (\bibinfo {year} {2004})},\ \Eprint
  {http://arxiv.org/abs/cond-mat/0305505} {arXiv:cond-mat/0305505} \BibitemShut
  {NoStop}%
\bibitem [{\citenamefont {{Wang}}\ and\ \citenamefont
  {{Senthil}}(2014)}]{2014PhRvB..89s5124W}%
  \BibitemOpen
  \bibfield  {author} {\bibinfo {author} {\bibfnamefont {C.}~\bibnamefont
  {{Wang}}}\ and\ \bibinfo {author} {\bibfnamefont {T.}~\bibnamefont
  {{Senthil}}},\ }\href {\doibase 10.1103/PhysRevB.89.195124} {\bibfield
  {journal} {\bibinfo  {journal} {\prb}\ }\textbf {\bibinfo {volume} {89}},\
  \bibinfo {eid} {195124} (\bibinfo {year} {2014})},\ \Eprint
  {http://arxiv.org/abs/1401.1142} {arXiv:1401.1142 [cond-mat.str-el]}
  \BibitemShut {NoStop}%
\bibitem [{\citenamefont {{Wang}}\ and\ \citenamefont
  {{Senthil}}(2016)}]{2016PhRvB..94x5107W}%
  \BibitemOpen
  \bibfield  {author} {\bibinfo {author} {\bibfnamefont {C.}~\bibnamefont
  {{Wang}}}\ and\ \bibinfo {author} {\bibfnamefont {T.}~\bibnamefont
  {{Senthil}}},\ }\href {\doibase 10.1103/PhysRevB.94.245107} {\bibfield
  {journal} {\bibinfo  {journal} {\prb}\ }\textbf {\bibinfo {volume} {94}},\
  \bibinfo {eid} {245107} (\bibinfo {year} {2016})},\ \Eprint
  {http://arxiv.org/abs/1604.06807} {arXiv:1604.06807 [cond-mat.str-el]}
  \BibitemShut {NoStop}%
\bibitem [{\citenamefont {{Sodemann}}\ \emph {et~al.}(2017)\citenamefont
  {{Sodemann}}, \citenamefont {{Kimchi}}, \citenamefont {{Wang}},\ and\
  \citenamefont {{Senthil}}}]{2017PhRvB..95h5135S}%
  \BibitemOpen
  \bibfield  {author} {\bibinfo {author} {\bibfnamefont {I.}~\bibnamefont
  {{Sodemann}}}, \bibinfo {author} {\bibfnamefont {I.}~\bibnamefont
  {{Kimchi}}}, \bibinfo {author} {\bibfnamefont {C.}~\bibnamefont {{Wang}}}, \
  and\ \bibinfo {author} {\bibfnamefont {T.}~\bibnamefont {{Senthil}}},\ }\href
  {\doibase 10.1103/PhysRevB.95.085135} {\bibfield  {journal} {\bibinfo
  {journal} {\prb}\ }\textbf {\bibinfo {volume} {95}},\ \bibinfo {eid} {085135}
  (\bibinfo {year} {2017})},\ \Eprint {http://arxiv.org/abs/1609.08616}
  {arXiv:1609.08616 [cond-mat.str-el]} \BibitemShut {NoStop}%
\bibitem [{\citenamefont {Wang}\ \emph {et~al.}(2017)\citenamefont {Wang},
  \citenamefont {Nahum}, \citenamefont {Metlitski}, \citenamefont {Xu},\ and\
  \citenamefont {Senthil}}]{Wang:2017txt}%
  \BibitemOpen
  \bibfield  {author} {\bibinfo {author} {\bibfnamefont {C.}~\bibnamefont
  {Wang}}, \bibinfo {author} {\bibfnamefont {A.}~\bibnamefont {Nahum}},
  \bibinfo {author} {\bibfnamefont {M.~A.}\ \bibnamefont {Metlitski}}, \bibinfo
  {author} {\bibfnamefont {C.}~\bibnamefont {Xu}}, \ and\ \bibinfo {author}
  {\bibfnamefont {T.}~\bibnamefont {Senthil}},\ }\href {\doibase
  10.1103/PhysRevX.7.031051} {\bibfield  {journal} {\bibinfo  {journal} {Phys.
  Rev. X}\ }\textbf {\bibinfo {volume} {7}},\ \bibinfo {pages} {031051}
  (\bibinfo {year} {2017})},\ \Eprint {http://arxiv.org/abs/1703.02426}
  {arXiv:1703.02426 [cond-mat.str-el]} \BibitemShut {NoStop}%
\bibitem [{\citenamefont {C\'ordova}\ and\ \citenamefont
  {Ohmori}(2019)}]{Cordova:2019bsd}%
  \BibitemOpen
  \bibfield  {author} {\bibinfo {author} {\bibfnamefont {C.}~\bibnamefont
  {C\'ordova}}\ and\ \bibinfo {author} {\bibfnamefont {K.}~\bibnamefont
  {Ohmori}},\ }\href@noop {} {\enquote {\bibinfo {title} {{Anomaly Obstructions
  to Symmetry Preserving Gapped Phases}},}\ } (\bibinfo {year} {2019}),\
  \Eprint {http://arxiv.org/abs/1910.04962} {arXiv:1910.04962 [hep-th]}
  \BibitemShut {NoStop}%
\bibitem [{\citenamefont {C\'ordova}\ and\ \citenamefont
  {Ohmori}(2020)}]{Cordova:2019jqi}%
  \BibitemOpen
  \bibfield  {author} {\bibinfo {author} {\bibfnamefont {C.}~\bibnamefont
  {C\'ordova}}\ and\ \bibinfo {author} {\bibfnamefont {K.}~\bibnamefont
  {Ohmori}},\ }\href {\doibase 10.1103/PhysRevD.102.025011} {\bibfield
  {journal} {\bibinfo  {journal} {Phys. Rev. D}\ }\textbf {\bibinfo {volume}
  {102}},\ \bibinfo {pages} {025011} (\bibinfo {year} {2020})},\ \Eprint
  {http://arxiv.org/abs/1912.13069} {arXiv:1912.13069 [hep-th]} \BibitemShut
  {NoStop}%
\bibitem [{\citenamefont {Apte}\ \emph {et~al.}(2023)\citenamefont {Apte},
  \citenamefont {Cordova},\ and\ \citenamefont {Lam}}]{Apte:2022xtu}%
  \BibitemOpen
  \bibfield  {author} {\bibinfo {author} {\bibfnamefont {A.}~\bibnamefont
  {Apte}}, \bibinfo {author} {\bibfnamefont {C.}~\bibnamefont {Cordova}}, \
  and\ \bibinfo {author} {\bibfnamefont {H.~T.}\ \bibnamefont {Lam}},\ }\href
  {\doibase 10.1103/PhysRevB.108.045134} {\bibfield  {journal} {\bibinfo
  {journal} {Phys. Rev. B}\ }\textbf {\bibinfo {volume} {108}},\ \bibinfo
  {pages} {045134} (\bibinfo {year} {2023})},\ \Eprint
  {http://arxiv.org/abs/2212.14605} {arXiv:2212.14605 [hep-th]} \BibitemShut
  {NoStop}%
\bibitem [{\citenamefont {Wang}\ and\ \citenamefont
  {Wen}(2023)}]{Wang:2013yta}%
  \BibitemOpen
  \bibfield  {author} {\bibinfo {author} {\bibfnamefont {J.}~\bibnamefont
  {Wang}}\ and\ \bibinfo {author} {\bibfnamefont {X.-G.}\ \bibnamefont {Wen}},\
  }\href {\doibase 10.1103/PhysRevB.107.014311} {\bibfield  {journal} {\bibinfo
   {journal} {Phys. Rev. B}\ }\textbf {\bibinfo {volume} {107}},\ \bibinfo
  {pages} {014311} (\bibinfo {year} {2023})},\ \Eprint
  {http://arxiv.org/abs/1307.7480} {arXiv:1307.7480 [hep-lat]} \BibitemShut
  {NoStop}%
\bibitem [{\citenamefont {Wang}\ and\ \citenamefont
  {Wen}(2018)}]{Wang:2018ugf}%
  \BibitemOpen
  \bibfield  {author} {\bibinfo {author} {\bibfnamefont {J.}~\bibnamefont
  {Wang}}\ and\ \bibinfo {author} {\bibfnamefont {X.-G.}\ \bibnamefont {Wen}},\
  }\href {\doibase 10.1103/PhysRevD.99.111501} {\bibfield  {journal} {\bibinfo
  {journal} {Phys. Rev. D}\ }\textbf {\bibinfo {volume} {99}},\ \bibinfo
  {pages} {111501} (\bibinfo {year} {2018})},\ \Eprint
  {http://arxiv.org/abs/1807.05998} {arXiv:1807.05998 [hep-lat]} \BibitemShut
  {NoStop}%
\bibitem [{\citenamefont {Chen}\ \emph {et~al.}(2014)\citenamefont {Chen},
  \citenamefont {Lu},\ and\ \citenamefont {Vishwanath}}]{CLV1301}%
  \BibitemOpen
  \bibfield  {author} {\bibinfo {author} {\bibfnamefont {X.}~\bibnamefont
  {Chen}}, \bibinfo {author} {\bibfnamefont {Y.-M.}\ \bibnamefont {Lu}}, \ and\
  \bibinfo {author} {\bibfnamefont {A.}~\bibnamefont {Vishwanath}},\ }\href
  {\doibase 10.1038/ncomms4507} {\bibfield  {journal} {\bibinfo  {journal} {Nat
  Commun}\ }\textbf {\bibinfo {volume} {5}},\ \bibinfo {pages} {4507} (\bibinfo
  {year} {2014})},\ \Eprint {http://arxiv.org/abs/1303.4301} {arXiv:1303.4301}
  \BibitemShut {NoStop}%
\bibitem [{\citenamefont {Li}\ \emph {et~al.}(2024)\citenamefont {Li},
  \citenamefont {Oshikawa},\ and\ \citenamefont {Zheng}}]{LOZ220403131}%
  \BibitemOpen
  \bibfield  {author} {\bibinfo {author} {\bibfnamefont {L.}~\bibnamefont
  {Li}}, \bibinfo {author} {\bibfnamefont {M.}~\bibnamefont {Oshikawa}}, \ and\
  \bibinfo {author} {\bibfnamefont {Y.}~\bibnamefont {Zheng}},\ }\href
  {\doibase 10.21468/SciPostPhys.17.1.013} {\bibfield  {journal} {\bibinfo
  {journal} {SciPost Phys.}\ }\textbf {\bibinfo {volume} {17}},\ \bibinfo
  {pages} {013} (\bibinfo {year} {2024})},\ \Eprint
  {http://arxiv.org/abs/2204.03131} {arXiv:2204.03131} \BibitemShut {NoStop}%
\bibitem [{\citenamefont {Levin}\ and\ \citenamefont {Gu}(2012)}]{LG1209}%
  \BibitemOpen
  \bibfield  {author} {\bibinfo {author} {\bibfnamefont {M.}~\bibnamefont
  {Levin}}\ and\ \bibinfo {author} {\bibfnamefont {Z.-C.}\ \bibnamefont {Gu}},\
  }\href {\doibase 10.1103/physrevb.86.115109} {\bibfield  {journal} {\bibinfo
  {journal} {Physical Review B}\ }\textbf {\bibinfo {volume} {86}},\ \bibinfo
  {pages} {115109} (\bibinfo {year} {2012})},\ \Eprint
  {http://arxiv.org/abs/1202.3120} {arXiv:1202.3120} \BibitemShut {NoStop}%
\bibitem [{\citenamefont {Ji}\ \emph {et~al.}(2020)\citenamefont {Ji},
  \citenamefont {Shao},\ and\ \citenamefont {Wen}}]{Ji:2019ugf}%
  \BibitemOpen
  \bibfield  {author} {\bibinfo {author} {\bibfnamefont {W.}~\bibnamefont
  {Ji}}, \bibinfo {author} {\bibfnamefont {S.-H.}\ \bibnamefont {Shao}}, \ and\
  \bibinfo {author} {\bibfnamefont {X.-G.}\ \bibnamefont {Wen}},\ }\href
  {\doibase 10.1103/PhysRevResearch.2.033317} {\bibfield  {journal} {\bibinfo
  {journal} {Phys. Rev. Res.}\ }\textbf {\bibinfo {volume} {2}},\ \bibinfo
  {pages} {033317} (\bibinfo {year} {2020})},\ \Eprint
  {http://arxiv.org/abs/1909.01425} {arXiv:1909.01425 [cond-mat.str-el]}
  \BibitemShut {NoStop}%
\bibitem [{\citenamefont {Thorngren}\ and\ \citenamefont
  {Wang}(2024{\natexlab{b}})}]{Thorngren:2021yso}%
  \BibitemOpen
  \bibfield  {author} {\bibinfo {author} {\bibfnamefont {R.}~\bibnamefont
  {Thorngren}}\ and\ \bibinfo {author} {\bibfnamefont {Y.}~\bibnamefont
  {Wang}},\ }\href {\doibase 10.1007/JHEP07(2024)051} {\bibfield  {journal}
  {\bibinfo  {journal} {JHEP}\ }\textbf {\bibinfo {volume} {07}},\ \bibinfo
  {pages} {051} (\bibinfo {year} {2024}{\natexlab{b}})},\ \Eprint
  {http://arxiv.org/abs/2106.12577} {arXiv:2106.12577 [hep-th]} \BibitemShut
  {NoStop}%
\bibitem [{\citenamefont {Vernier}\ \emph {et~al.}(2019)\citenamefont
  {Vernier}, \citenamefont {O'Brien},\ and\ \citenamefont
  {Fendley}}]{Vernier:2018han}%
  \BibitemOpen
  \bibfield  {author} {\bibinfo {author} {\bibfnamefont {E.}~\bibnamefont
  {Vernier}}, \bibinfo {author} {\bibfnamefont {E.}~\bibnamefont {O'Brien}}, \
  and\ \bibinfo {author} {\bibfnamefont {P.}~\bibnamefont {Fendley}},\ }\href
  {\doibase 10.1088/1742-5468/ab11c0} {\bibfield  {journal} {\bibinfo
  {journal} {J. Stat. Mech.}\ }\textbf {\bibinfo {volume} {1904}},\ \bibinfo
  {pages} {043107} (\bibinfo {year} {2019})},\ \Eprint
  {http://arxiv.org/abs/1812.09091} {arXiv:1812.09091 [cond-mat.stat-mech]}
  \BibitemShut {NoStop}%
\bibitem [{\citenamefont {Miao}(2021)}]{Miao:2021pyh}%
  \BibitemOpen
  \bibfield  {author} {\bibinfo {author} {\bibfnamefont {Y.}~\bibnamefont
  {Miao}},\ }\href {\doibase 10.21468/SciPostPhys.11.3.066} {\bibfield
  {journal} {\bibinfo  {journal} {SciPost Phys.}\ }\textbf {\bibinfo {volume}
  {11}},\ \bibinfo {pages} {066} (\bibinfo {year} {2021})},\ \Eprint
  {http://arxiv.org/abs/2103.14569} {arXiv:2103.14569 [cond-mat.stat-mech]}
  \BibitemShut {NoStop}%
\bibitem [{\citenamefont {{Popkov}}\ \emph {et~al.}(2024)\citenamefont
  {{Popkov}}, \citenamefont {{Zhang}}, \citenamefont {{G\"ohmann}},\ and\
  \citenamefont {{Kl\"umper}}}]{PZG230314056}%
  \BibitemOpen
  \bibfield  {author} {\bibinfo {author} {\bibfnamefont {V.}~\bibnamefont
  {{Popkov}}}, \bibinfo {author} {\bibfnamefont {X.}~\bibnamefont {{Zhang}}},
  \bibinfo {author} {\bibfnamefont {F.}~\bibnamefont {{G\"ohmann}}}, \ and\
  \bibinfo {author} {\bibfnamefont {A.}~\bibnamefont {{Kl\"umper}}},\ }\href
  {\doibase 10.1103/PhysRevLett.132.220404} {\bibfield  {journal} {\bibinfo
  {journal} {Phys. Rev. Lett.}\ }\textbf {\bibinfo {volume} {132}},\ \bibinfo
  {pages} {220404} (\bibinfo {year} {2024})},\ \Eprint
  {http://arxiv.org/abs/2303.14056} {arXiv:2303.14056 [quant-ph]} \BibitemShut
  {NoStop}%
\bibitem [{\citenamefont {Pace}\ \emph
  {et~al.}(2024{\natexlab{a}})\citenamefont {Pace}, \citenamefont
  {Chatterjee},\ and\ \citenamefont {Shao}}]{Pace:2024oys}%
  \BibitemOpen
  \bibfield  {author} {\bibinfo {author} {\bibfnamefont {S.~D.}\ \bibnamefont
  {Pace}}, \bibinfo {author} {\bibfnamefont {A.}~\bibnamefont {Chatterjee}}, \
  and\ \bibinfo {author} {\bibfnamefont {S.-H.}\ \bibnamefont {Shao}},\
  }\href@noop {} {\  (\bibinfo {year} {2024}{\natexlab{a}})},\ \Eprint
  {http://arxiv.org/abs/2412.18606} {arXiv:2412.18606 [cond-mat.str-el]}
  \BibitemShut {NoStop}%
\bibitem [{\citenamefont {Seiberg}\ \emph {et~al.}(2024)\citenamefont
  {Seiberg}, \citenamefont {Seifnashri},\ and\ \citenamefont
  {Shao}}]{Seiberg:2024gek}%
  \BibitemOpen
  \bibfield  {author} {\bibinfo {author} {\bibfnamefont {N.}~\bibnamefont
  {Seiberg}}, \bibinfo {author} {\bibfnamefont {S.}~\bibnamefont {Seifnashri}},
  \ and\ \bibinfo {author} {\bibfnamefont {S.-H.}\ \bibnamefont {Shao}},\
  }\href {\doibase 10.21468/SciPostPhys.16.6.154} {\bibfield  {journal}
  {\bibinfo  {journal} {SciPost Phys.}\ }\textbf {\bibinfo {volume} {16}},\
  \bibinfo {pages} {154} (\bibinfo {year} {2024})},\ \Eprint
  {http://arxiv.org/abs/2401.12281} {arXiv:2401.12281 [cond-mat.str-el]}
  \BibitemShut {NoStop}%
\bibitem [{\citenamefont {Gorantla}\ \emph {et~al.}(2024)\citenamefont
  {Gorantla}, \citenamefont {Shao},\ and\ \citenamefont
  {Tantivasadakarn}}]{Gorantla:2024ocs}%
  \BibitemOpen
  \bibfield  {author} {\bibinfo {author} {\bibfnamefont {P.}~\bibnamefont
  {Gorantla}}, \bibinfo {author} {\bibfnamefont {S.-H.}\ \bibnamefont {Shao}},
  \ and\ \bibinfo {author} {\bibfnamefont {N.}~\bibnamefont
  {Tantivasadakarn}},\ }\href@noop {} {\enquote {\bibinfo {title} {{Tensor
  networks for non-invertible symmetries in 3+1d and beyond}},}\ } (\bibinfo
  {year} {2024}),\ \Eprint {http://arxiv.org/abs/2406.12978} {arXiv:2406.12978
  [quant-ph]} \BibitemShut {NoStop}%
\bibitem [{\citenamefont {Pace}\ \emph
  {et~al.}(2024{\natexlab{b}})\citenamefont {Pace}, \citenamefont {Delfino},
  \citenamefont {Lam},\ and\ \citenamefont {Aksoy}}]{PGLA240612962}%
  \BibitemOpen
  \bibfield  {author} {\bibinfo {author} {\bibfnamefont {S.~D.}\ \bibnamefont
  {Pace}}, \bibinfo {author} {\bibfnamefont {G.}~\bibnamefont {Delfino}},
  \bibinfo {author} {\bibfnamefont {H.~T.}\ \bibnamefont {Lam}}, \ and\
  \bibinfo {author} {\bibfnamefont {{\"O}.~M.}\ \bibnamefont {Aksoy}},\
  }\href@noop {} {\enquote {\bibinfo {title} {{Gauging modulated symmetries:
  Kramers-Wannier dualities and non-invertible reflections}},}\ } (\bibinfo
  {year} {2024}{\natexlab{b}}),\ \Eprint {http://arxiv.org/abs/2406.12962}
  {arXiv:2406.12962 [cond-mat.str-el]} \BibitemShut {NoStop}%
\bibitem [{\citenamefont {Metlitski}\ and\ \citenamefont
  {Thorngren}(2018)}]{Metlitski:2017fmd}%
  \BibitemOpen
  \bibfield  {author} {\bibinfo {author} {\bibfnamefont {M.~A.}\ \bibnamefont
  {Metlitski}}\ and\ \bibinfo {author} {\bibfnamefont {R.}~\bibnamefont
  {Thorngren}},\ }\href {\doibase 10.1103/PhysRevB.98.085140} {\bibfield
  {journal} {\bibinfo  {journal} {Phys. Rev. B}\ }\textbf {\bibinfo {volume}
  {98}},\ \bibinfo {pages} {085140} (\bibinfo {year} {2018})},\ \Eprint
  {http://arxiv.org/abs/1707.07686} {arXiv:1707.07686 [cond-mat.str-el]}
  \BibitemShut {NoStop}%
\end{thebibliography}%

\onecolumngrid
\appendix

\renewcommand{\theequation}{\thesection.\arabic{equation}}
\makeatletter\@addtoreset{equation}{section}\makeatother

\renewcommand{\QVA}{iG_1}

\section{\texorpdfstring{Continuum limits of $[\QV,\QA]$ and of the Onsager algebra}{Continuum limits of [QV,QA] and of the Onsager algebra}}\label{app:commutator}

In the main text, we claim that the quantized lattice vector and axial charges $\QV$ and $\QA$, given by Eqs.~\eqref{eq:QVdef} and~\eqref{eq:QAdef} respectively, commute in the continuum limit 
 of the staggered fermion Hamiltonian.  
In this appendix, we prove this statement. 
More precisely, we will show that the matrix elements of ${[\QV,\QA]}$ between any two low-energy states vanish in the continuum limit ${L\to \infty}$. 

Low-energy states of the staggered fermion Hamiltonian model are most naturally understood in momentum space, where they reside near momenta ${k=0}$ and ${k = L/2}$. Recall that in terms of the momentum space operators ${\gamma_k = \frac{1}{\sqrt{L}}\sum_{j=1}^L e^{-\frac{2\pi i k j}L} c_j}$, the Hamiltonian is re-written as
\ie 
H = \sum_{-\frac L2<  k\leq \frac L2} 2 \sin\frac{2\pi k}{L}\  \gamma_k^\dagger \gamma_k^{\,}.
\fe 
When the fermions obey periodic boundary conditions, there are zero modes at ${k=0}$ and ${k=L/2}$, which cause the ground states ${\ket{\Om;J}}$ to be 4-fold degenerate (${J=1,2,3,4}$). 
For anti-periodic boundary conditions, ${k=0}$ and ${k=L/2}$ are not allowed values of momenta, and there is a single ground state ${\ket{\Om;J}}$ with ${J = 1}$. For both periodic and anti-periodic boundary conditions, the ground state(s) satisfy
\begin{equation}\label{GSdef}
    \ga_k^\dagger\, \ga_k^{\,} \ket{\Om;J} = 
    \begin{cases}
        \ket{\Om;J}\qquad &-\frac L2 < k < 0,\\
        0\qquad &\ \ \ 0<k<\frac L2,
    \end{cases}
\end{equation}
and the low-energy states are created by acting $\ga_k^{\,}$ and $\ga_k^\dagger$ on ${\ket{\Om;J}}$ with $k$ near $0$ or ${L/2}$. 
Recall from the main text that the commutator $[\QV,\QA]$ gives the operator $G_1$ in the Onsager algebra:
\ie 
\QVA = 
[\QV,\QA] = -\sum_{j=1}^L \left(c_j^{\,} c_{j+1}^{\,} 
+ c_j^\dagger 
c_{j+1}^\dagger
\right) \, ,
\fe 
which in momentum space is
\ie \label{eq:QVQAcomk}
\QVA
= \sum_{k}\left(e^{\frac{2\pi i}{L}k}\ \gamma_{-k}^\dagger \gamma_{k}^\dagger
-
e^{-\frac{2\pi i}{L}k}\ \gamma_k^{\,} \gamma_{-k}^{\,}  
\right)\, .
\fe 

Consider two arbitrary low-energy states 
${\ket{\phi_1} = \gamma_{k_1}^\dagger\dots \gamma_{k_n}^\dagger \gamma_{q_1}^{\,} \dots \gamma_{q_m}^{\,} \ket{\Omega,J} }$ and 
${\ket{\phi_2} = \gamma_{k'_1}^\dagger\dots \gamma_{k'_{n'}}^\dagger \gamma_{q'_1}^{\,} \dots \gamma_{q'_{m'}}^{\,} \ket{\Omega,J'} }$ and the related matrix element of the commutator
\ie \label{matrixEle0}
\bra{\phi_2}\QVA \ket{\phi_1} &= 
\bra{\Omega,J'} \gamma_{q'_{m'}}^\dagger\dots 
\gamma_{q'_1}^{\dagger} \gamma_{k'_{n'}}^{\,} \dots \gamma_{k'_1}^{\,}
~\QVA ~
\gamma_{k_1}^\dagger\dots \gamma_{k_n}^\dagger \gamma_{q_1}^{\,} \dots \gamma_{q_m}^{\,} \ket{\Omega,J}.
\fe

In the continuum limit, this matrix element simplifies. Indeed, using~\eqref{eq:QVQAcomk}, it follows that 
\ie \label{eq:G1,gamma_k}
{[\QVA, \gamma_{k}^\dagger]
=
- 2i \sin\frac{2 \pi k}{L} 
\gamma_{-k}^{\,}},
\hspace{40pt}
{[\QVA, \gamma_{k}^{\,}]
=
2i \sin\frac{2 \pi k}{L} 
\gamma_{-k}^\dagger}\,.
\fe
For $k$ close to 0 or ${L/2}$, these commutators vanish in the ${L\to\infty}$ limit. Therefore, the commutator $\QVA$ commutes with all low-energy operators in the continuum limit. Using this, we can rewrite the matrix elements~\eqref{matrixEle0} in the continuum limit as
\ie \label{matrixElecont}
\lim_{L\to\infty}\, \bra{\phi_2}\QVA \ket{\phi_1} &= 
\bra{\Omega,J'} \gamma_{q'_{m'}}^\dagger\dots 
\gamma_{q'_1}^{\dagger} \gamma_{k'_{n'}}^{\,} \dots \gamma_{k'_1}^{\,}\,
\gamma_{k_1}^\dagger\dots \gamma_{k_n}^\dagger \gamma_{q_1}^{\,} \dots \gamma_{q_m}^{\,} ~\QVA\,\ket{\Omega,J}.
\fe 
However, using Eqs.~\eqref{GSdef} and~\eqref{eq:QVQAcomk}, the commutator $\QVA$ satisfies ${\QVA \ket{\Omega,J}= 0}$. 
This is because for any ${k\neq 0,L/2}$, either $\ga_k^{\,}$ or $\ga_{-k}^{\,}$ annihilates $\ket{\Om,J}$, and so ${\ga_k^{\,}\ga_{-k}^{\,}}$ necessarily annihilates $\ket{\Om,J}$. 
Furthermore, this is true for both periodic and anti-periodic boundary conditions. 
Indeed, for the zero modes ${k=0, L/2}$ occuring from periodic boundary conditions, we have $\ga_k^{\,}\ga_{-k}^{\,}=\ga_k^{\,}\ga_{k}^{\,} = 0$ and so these values of $k$ do not contribute to $\QVA$. 
Following this argument, the matrix element~\eqref{matrixElecont} simplifies to
\ie \label{matrixElecont2}
\lim_{L\to\infty}\, \bra{\phi_2}\QVA \ket{\phi_1} &= 
0.
\fe 
Since $\ket{\phi_1}$ and $\ket{\phi_2}$ are arbitrary low-energy states, we conclude that all matrix elements of ${\QVA\equiv [\QV,\QA]}$ involving low-energy states vanish in the continuum limit, and $\QV$ and $\QA$ commute in the continuum limit.

There is a more general statement one can make about the fate of the Onsager algebra in the continuum limit. Not only $G_1$, but $G_n$ for all $n\ll L$ vanish as operators, when restricted to the low-energy states. Recall that
\ie 
i G_n = -\frac{1}2\sum_j( a_j a_{j+n} - b_j b_{j+n})
= -\sum_j\left( c_j^{\,} c_{j+n}^{\,} + c_j^\dagger c_{j+n}^\dagger\right )
\,.
\fe 
Then it follows 
from its momentum space expression
that the generalization of \eqref{eq:G1,gamma_k} is 
\ie \label{eq:Gn,gamma_k}
{[iG_n, \gamma_{k}^\dagger]
=
- 2i \sin\frac{2 \pi nk}{L} 
\gamma_{-k}^{\,}} \, ,
\hspace{40pt}
{[iG_n, \gamma_{k}^{\,}]
=
2i \sin\frac{2 \pi nk}{L} 
\gamma_{-k}^\dagger}\,.
\fe
Again, these commutators vanish in the ${L\to\infty}$ limit for $k$ close to 0 or $L/2$. The remainder of the argument demonstrating the vanishing of $\QVA$ in the continuum limit goes through identically for all $iG_n$ (with $n\ll L$).

\section{\texorpdfstring{Deformations preserving both $\QV$ and $\QA$}{Deformations preserving both QV and QA}}
\label{app:symdef}

In the main text, we discussed how the lattice vector and axial symmetries are anomalous in the sense that there does not exist a non-degenerated gapped Hamiltonian with U(1)$^\mathrm{V}$ and U(1)$^\mathrm{A}$ symmetries. In fact, we claimed that all local Hamiltonians that are U(1)$^\mathrm{V}$ and U(1)$^\mathrm{A}$ symmetric take the form
\ie \label{genHam}
H &= \sum_{n=1}^N \left( -i\, g_n
\sum_{j=1}^L
(c_j^\dagger c_{j+n}^{\,} + c_j^{\,} c_{j+n}^\dagger ) \right),
\fe 
with ${\lim_{L\to \infty} N/L = 0}$.
This is a gapless Hamiltonian for all couplings $g_n$ with dispersion ${\om_k = 2\sum_{n=1}^N g_n \sin(2\pi nk/L)}$. In this appendix, we prove this statement. In particular, we will now prove that all local deformations of the staggered fermion Hamiltonian that commute with the U(1)$^\mathrm{V}$ symmetry operator ${e^{i\varphi\QV}}$ and the ${\Z_4^\mathrm{A}\subset \text{U(1)}^{\mathrm{A}}}$ symmetry operator ${e^{-i\frac\pi2\QA}}$ are quadratic and of the type~\eqref{genHam}, which has the full U(1)$^\mathrm{A}$ symmetry.

We start by considering the product of symmetry operators ${e^{-i\frac\pi2\QA} e^{i\frac\pi2\QV}}$, which satisfies ${e^{-i\frac\pi2\QA} e^{i\frac\pi2\QV} = T_b^{\,} T_a^{-1}}$. Indeed, this identity can be seen by noting that $e^{-i\frac\pi2\QA} e^{i\frac\pi2\QV}$ act on $a_j$ and $b_j$ as
\ie 
e^{-i\frac\pi2\QA} e^{i\frac\pi2\QV} a_j e^{-i\frac\pi2\QV} e^{i\frac\pi2\QA} 
&= e^{-i\frac\pi2\QA} b_j e^{i\frac\pi2\QA} 
= a_{j-1} \, ,\\
e^{-i\frac\pi2\QA} e^{i\frac\pi2\QV} b_j e^{-i\frac\pi2\QV} e^{i\frac\pi2\QA}
&= e^{-i\frac\pi2\QA} (-a_j) e^{i\frac\pi2\QA} 
= b_{j+1},
\fe
which is the same action as ${T_b^{\,} T_a^{-1}}$. 
Then, because $a_j$ and $b_j$ form a faithful basis for the algebra of operators acting on this tensor product Hilbert space, the operator identity ${e^{-i\frac\pi2\QA} e^{i\frac\pi2\QV} = T_b^{\,} T_a^{-1}}$ holds. 
Imposing the symmetry generated by ${T_b^{\,} T_a^{-1}}$ forbids deformations to the staggered fermion Hamiltonian that couple the $a$ and $b$ Majorana fermions. 
This is because repeated actions of ${T_b^{\,} T_a^{-1}}$ increase the distance between the $a$ and $b$ Majorana operators, so that any local product of them will become increasingly non-local. 
Therefore, by imposing that all deformations commute with ${T_b^{\,} T_a^{-1}}$, we need only consider deformations whose individual terms consist of only the $a_j$ Majorana fermions or only the $b_j$ Majorana fermions. 
Furthermore, by locality of the Hamiltonian, this implies that $T_a$ and $T_b$ individually commute as well, and therefore lattice translations ${T=T_a T_b}$ too.

Having found the effect of requiring ${T_b^{\,} T_a^{-1}}$ invariance, we next require all deformations to commute with ${e^{i\varphi\QV}}$. Let us first consider the ${T_b^{\,} T_a^{-1}}$-symmetric Majorana bi-linears
\ie \label{eq:2bdy}
O_2^a(j,j') = a_j a_{j'},
\hspace{40pt}
 O_2^b(j,j') = b_jb_{j'} ,
\fe 
where we assume ${j<j'}$ without loss of generality. 
An infinitesimal $\QV$ action transforms these terms by
\ie 
\del O_2^a(j,j') &= \varphi (b_j a_{j'} + a_j b_{j'} ) + \cO(\varphi^2), \\
\del O_2^b(j,j') &= -\varphi (a_j b_{j'} + b_j a_{j'} ) + \cO(\varphi^2).
\fe 
Therefore, the only combinations of these bi-linears that are U(1)$^\mathrm{V}$ symmetric take the form 
\ie 
O_2^a(j,j') + O_2^b(j,j') &= a_ja_{j'} + b_jb_{j'} 
= 2(\,c_j^\dagger c_{j'}^{\,} + c_j^{\,} c_{j'}^\dagger\,)
\, ,
\fe 
and the most general $T_b^{\,} T_a^{-1}$ and U(1)$^\mathrm{V}$ symmetric, local deformation using fermion bi-linears is
\begin{equation}\label{eq:dhn}
    \del H = -\frac i2 \sum_{j=1}^L \sum_{n=1}^{N} g_{n} (a_ja_{j+n} + b_jb_{j+n} ),
\end{equation}
where ${\lim_{L\to \infty} N/L = 0}$. Note that the original staggered fermion Hamiltonian~\eqref{eq:Dirac} is of such type for which ${n=1}$. Furthermore, because $\del H$ commutes with $T_b$ and $\QV$, in light of \eqref{eq:QVQA}, it also commutes with $\QA$ too.

Having considered symmetric bi-linear operators, we now consider $T_b^{\,} T_a^{-1}$-symmetric $2n$-body terms with ${n\geq 2}$, which take the form
\ie 
O_{2n}^a(j_1,\dots, j_{2n}) &= \prod_{\ell=1}^{2n} a_{j_\ell} ,
\hspace{40pt}
O_{2n}^b(j_1,\dots, j_{2n}) 
= \prod_{\ell=1}^{2n} b_{j_\ell} \, .
\fe 
An infinitesimal $\QV$ action transforms these operators by
\ie 
\del O_{2n}^a(\{j_\ell\}) &= \varphi \sum_{\ell'=1}^{2n} a_{j_1}\cdots b_{j_{\ell'}} \cdots a_{j_{2n}} +\cO(\varphi^2), \\
\del O_{2n}^b(\{j_\ell\}) &= -\varphi \sum_{\ell'=1}^{2n} b_{j_1}\cdots a_{j_{\ell'}} \cdots b_{j_{2n}} +\cO(\varphi^2) \, .
\fe 
Because each term in ${\del O_{2n}^a}$ consists of ${2n-1}$ $a_j$ operators and $1$ $b_j$ operator, while each term in ${\del O_{2n}^b}$ contains ${2n-1}$ $b_j$ operators and $1$ $a_j$ operator, ${\del O_{2n}^a}$ and ${\del O_{2n}^b}$ are linearly independent when ${n\geq 2}$. Therefore, there is no linear combination of ${O_{2n}^a(j_1,\dots, j_{2n})}$ and ${O_{2n}^b(j_1,\dots, j_{2n})}$ for ${n\geq 2}$ that is symmetric under the entire U(1)$^\mathrm{V}$ symmetry, and the only allowed deformations of the staggered fermion model~\eqref{eq:Maj} are quadratic terms of the form~\eqref{eq:dhn}.

\section{\texorpdfstring{$e^{i\lm \tQA}$}{exp(i tQA)} is not a locality-preserving unitary}\label{actionApp}

The commutator between the unquantized axial charge $\tQA$ and the fermion creation operator in momentum space is
\ie
[\tQA , \gamma_k^\dagger  ] =\cos {\frac{2\pi k}{L}} \gamma_k^\dagger \,.
\fe
Consider the exponentiated operator $e^{ i  \lambda \tQA}$. 
Since $\tQA$ is not quantized, $\lambda$ is $\mathbb{R}$-valued. 
The exponentiated operator has the following commutation relation with the fermion creation operator
\ie
[e^{ i  \lambda \tQA}, \gamma_k^\dagger  ]
=e^{i  \lambda\cos {\frac{2\pi k}{L}}} \gamma_k^\dagger \,.
\fe
To find the action of $e^{ i  \lambda \tQA}$ in position space, we perform the Fourier transformation on both sides: 
\ie
[e^{ i  \lambda \tQA}, c_j^\dagger  ]
= {\frac1{L}}
\sum_{j'=1}^L \sum_k  
e^{i  \lambda\cos(\frac{2\pi k} L)} 
e^{ -\frac{2\pi i  (j-j') k}L} 
c_{j'}^\dagger
\equiv \sum_{j'=1}^L A_\lambda(j -j') c_{j'}^\dagger
\fe
where
\ie
A_\lambda(  j -j' ) = \frac1{L} 
\sum_k  
e^{i  \lambda\cos \frac{2\pi k}L} 
e^{ -\frac{2\pi i  (j-j') k}L}  \,.
\fe

\begin{figure}
    \centering
\includegraphics[width=0.55\linewidth]{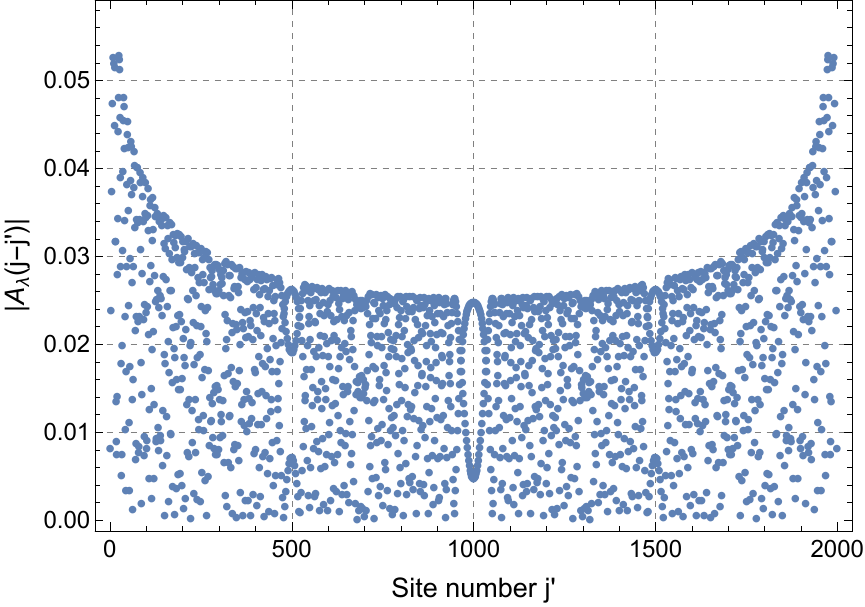}
    \caption{Plot of the absolute value $|A_\lm(j-j')|$ for ${L = 2000}$ and ${j=\lm = L/2}$.}
    \label{fig:plot}
\end{figure}

In the large $L$ limit, we can approximate the sum by a continuous integral of ${p \equiv 2\pi k / L}$:
\ie
A_\lambda (j-j')\sim \frac{1}{ 2\pi} \int_{-\pi}^\pi dp \,
e^{ - i  p ( j-j')  + i  \lambda \cos p}
= i ^{j-j'} J_{j-j'}(\lambda) \,.
\fe
(Note that this equality only holds if ${j-j'}$ is an integer.)
If we further take the large ${|j-j'|}$ limit, it becomes
\ie
A_\lambda( j - j' ) \sim 
i ^{j-j'} 
{ \sqrt{\frac{1}{2 \pi |j-j'|}}} 
\left(
\frac{e \lambda}{2 |j-j'|}
\right)^{|j-j'|} \,.
\fe
For a fixed $\lambda$, we see that ${A_\lambda(j -j')}$ decays as ${\exp ( - |j-j'| \log |j-j'|)}$ in the large ${j-j'}$ limit. 
Hence, while the action of $e^{i  \lambda \tQA}$ is not ultra-local, it is local for a fixed $\lambda$.

However, since the symmetry generated by $\tQA$ is $\mathbb{R}$ rather than U(1), $\lambda$ can be any real number. 
For a finite lattice (or a fixed range of $j-j'$), one can always take a sufficiently large $\lambda$ so that the action of $e^{i  \lambda\tQA}$ becomes highly non-local. For example, we plot the absolute value of $A_{\lm}(j-j')$ for ${\lm = j = L/2}$ in Fig.~\ref{fig:plot}, showing that it is not localized near ${j = L/2}$ and $e^{i \lm \tQA}$ for these parameters is not a locality preserving unitary operator.

\end{document}